\documentclass[fleqn,usenatbib]{mnras}

\usepackage[T1]{fontenc}

\DeclareRobustCommand{\VAN}[3]{#2}
\let\VANthebibliography\thebibliography
\def\thebibliography{\DeclareRobustCommand{\VAN}[3]{##3}\VANthebibliography}

\usepackage{graphicx}	
\usepackage{amsmath}	
\usepackage{amssymb}	
\usepackage{subcaption}
\usepackage{caption}
\usepackage{newtxtext,newtxmath}

\newcommand{\red}[1]{\textcolor{black}{#1}}

\title[AbacusHOD]{\textsc{AbacusHOD}: A highly efficient extended multi-tracer HOD framework and its application to BOSS and eBOSS data}

\author[S. Yuan et al.]{
Sihan Yuan,$^{1, 2}$\thanks{E-mail: sihany@stanford.edu}
Lehman H. Garrison, $^{3}$
Boryana Hadzhiyska, $^{1}$
Sownak Bose, $^{1,4}$
and Daniel J. Eisenstein $^{1}$
\\
$^{1}$Center for Astrophysics | Harvard \& Smithsonian, 60 Garden St., Cambridge, MA 02138, USA\\
$^{2}$Kavli Institute for Particle Astrophysics and Cosmology, Stanford University, Stanford, CA 94305, USA\\
$^{3}$Center for Computational Astrophysics, Flatiron Institute, 162 Fifth Avenue, New York, NY 10010, USA\\
$^{4}$Institute for Computational Cosmology, Department of Physics, Durham University, Durham DH1 3LE, UK
}

\date{Accepted XXX. Received YYY; in original form ZZZ}

\pubyear{2021}
\newcommand{\ahod}{\textsc{AbacusHOD}}

\begin{document}
\label{firstpage}
\pagerange{\pageref{firstpage}--\pageref{lastpage}}
\maketitle

\begin{abstract}
\red{
We introduce the \ahod\ model and present two applications of \ahod\ and the \textsc{AbacusSummit} simulations to observations. \ahod\ is an HOD framework written in \textsc{Python} that is particle-based, multi-tracer, highly generalized, and highly efficient. It is designed specifically with multi-tracer/cosmology analyses for next generation large-scale structure surveys in mind, and takes advantage of the volume and precision offered by the new state-of-the-art \textsc{AbacusSummit} cosmological simulations. The model is also highly customizable and should be broadly applicable to any upcoming surveys and a diverse range of cosmological analyses. In this paper, we demonstrate the capabilities of the \ahod\ framework through two example applications. The first example demonstrates the high efficiency and the large HOD extension feature set through an analysis of full-shape redshift-space clustering of BOSS galaxies at intermediate to small scales ($<30h^{-1}$Mpc), assessing the necessity of introducing secondary galaxy biases (assembly bias). We find strong evidence for using halo environment instead of concentration to trace secondary galaxy bias, a result which also leads to a moderate reduction to the ``lensing is low'' tension. The second example demonstrates the multi-tracer capabilities of the \ahod\ package through an analysis of the extended Baryon Oscillation Spectroscopic
Survey (eBOSS) cross-correlation measurements between three different galaxy tracers, LRGs, ELGs, and QSOs. We expect the \ahod\ framework, in combination with the \textsc{AbacusSummit} simulation suite, to play an important role in a simulation-based analysis of the up-coming Dark Energy Spectroscopic Instrument (DESI) datasets. }

\end{abstract}



\section{Introduction}
In the standard framework of structure formation in a $\Lambda$CDM universe, galaxies are predicted to form and evolve in dark matter halos \citep{1978White}.
To extract cosmological information and understand galaxy formation from  observed galaxy clustering statistics, it is critical to correctly model the connection between galaxies and their underlying dark matter halos. The most popular and efficient model of the galaxy-halo connection for cosmological studies is the Halo Occupation Distribution model \citep[HOD; e.g.][]{2000Peacock, 2001Scoccimarro, 2002Berlind, 2005Zheng, 2007bZheng}. The HOD is an empirical model that makes the assumption that all galaxies live inside dark matter halos, and links galaxy occupation to specific halo properties. The most popular form of the HOD assumes that galaxy occupation is determined {\it solely} by halo mass, an assumption that rests on the long-standing and widely accepted theoretical prediction that halo mass is the attribute that most strongly correlates with the halo abundance and halo clustering as well as the properties of the galaxies residing in it \citep[][]{1978White, 1984Blumenthal}. 

However, there is mounting evidence that the mass-only HOD model is insufficient in accurately capturing the observed galaxy clustering on moderate to small scales (around and less than $10 h^{-1}$Mpc). A series of studies employing hydrodynamical simulations and semi-analytic models have found clear evidence that galaxy occupation correlates with secondary halo properties beyond just halo mass \citep[e.g.][]{2006Zhu, 2018Artale, 2018Zehavi, 2019Bose, 2019Contreras, 2020Hadzhiyska, 2020Xu}. This phenomenon is commonly known as \textit{Galaxy Assembly Bias}, or just assembly bias hereafter. In \citet{2021Yuan}, we present observational evidence for galaxy assembly bias by analyzing the full-shape redshift-space clustering of BOSS CMASS galaxies \citep[][]{2011Eisenstein, 2013Dawson}. We found that by including assembly bias and a secondary dependence on the local environment, the HOD model makes a significantly better prediction of the observed redshift-space clustering. We also found that the predicted galaxy-galaxy lensing signal also becomes significantly more consistent with data, thus potentially alleviating the ``lensing is low'' tension, where the observed lensing signal was consistently lower than model predictions by $30-40\%$  \citep[][]{2017Leauthaud, 2019bLange}. All these studies combine to refute the mass-only ansatz of the basic HOD model, demanding a set of robust physically-motivated extensions to improve the HOD's predictive power. 

As shown in \citet{2021Yuan}, a key challenge with extended HODs is computational efficiency. The secondary dependencies significantly increase the complexity of the model. Additionally, in order to produce more physical galaxy distributions, we adopt a particle-based approach, where we directly link galaxies to the dark matter particles, as opposed to estimating galaxy positions according to analytical models. However, due to the large number of particles ($>100$ times that of the halos) in a high-resolution simulation box, a particle-based approach also significantly increases the computational cost. The combination of using particles and introducing secondary dependencies can make the HOD code too slow for comprehensive parameter space explorations. Indeed, a shortcoming of the \citet{2021Yuan} analysis is that instead of sampling the full extended HOD posterior space, we opted for a much cheaper global optimization routine, which could potentially miss interesting regions of parameter space, particularly in high dimension spaces. Moreover, future cosmological applications of the extended HODs will likely require HOD sampling at a range of different cosmologies, further increasing the computational cost. Thus, performance is of great priority for a robust extended HOD code. 

With the advent of a new generation of cosmological surveys with much greater depth, such as the extended Baryon Oscillation Spectroscopic Survey \citep[eBOSS;][]{2016Dawson} and the Dark Energy Spectroscopic Instrument \citep[DESI;][]{2016DESI}, there arises a new opportunity to simultaneously utilize multiple galaxy tracer types to probe structure, the so-called multi-tracer analysis \citep[e.g.][]{2020Alam, 2021Zhao}. There are three types of galaxies that are most relevant in current and upcoming cosmological surveys: luminous red galaxies (LRGs), which tend to be massive, spheroidal, and quenched; emission line galaxies (ELGs), which tend to be less massive, disk-like, and star-forming; and quasi stellar objects (QSOs), whose emissions are dominated by their active galactic nuiclei (AGNs). Multi-tracer studies can not only bring additional statistical power to cosmology studies, but also leverage the potential difference in the clustering of different galaxy types to constrain the physics of galaxy formation. To enable such multi-tracer analyses, it is extremely helpful to devise a multi-tracer HOD model, where we simultaneously assign multiple types of galaxies to each halo. 

In this paper, we introduce the \ahod\ framework and the accompanying \textsc{Python} code module. This HOD framework is extended, efficient, and multi-tracer. It is developed in conjunction with the state-of-the-art \textsc{AbacusSummit} simulations \citep{2021Maksimova} and designed specifically with DESI in mind. However, the model framework is also broadly applicable to other simulations and surveys. This paper also presents two applications of the \ahod\ framework and the \textsc{AbacusSummit} simulations to observational data, demonstrating its effectiveness in modeling observed galaxy clustering down to highly non-linear scales. 

We should also mention that there have been several other important HOD analysis frameworks aimed at efficiently deriving HOD fits. One is the so-called emulator approach \citep[e.g.][]{2019DeRose, 2019Zhai, 2019Wibking, 2019bWibking}, where a surrogate model is used to approximate the complex clustering predictions. This approach saves computation by training the surrogate model on a modest number of HOD realizations. However, the success of an emulator model relies delicately on the choice of the surrogate model, while the accuracy often falls off drastically outside the training range. Another approach that avoids emulation is often referred to as tabulation (first described in \citet{2016Zheng}, a popular example is \textsc{TabCorr}\footnote{\url{https://github.com/johannesulf/TabCorr}}), which focuses on minimizing the cost of re-computing clustering measurements. These approaches pre-tabulate the halo pair counts and then compute each HOD evaluation as a re-weighted sum of the tabulated halo pair counts. While tabulation can make HOD evaluations quite fast, it is limited by the type of statistics and the binning beforehand. HOD extensions and introducing particles also significantly increase the complexity in tabulated approaches. The \ahod\ framework aims to bypass the limits of emulators and tabulation by directly optimizing HOD evaluation and clustering calculation, ensuring maximum flexibility in the HOD model itself. 

The paper is outlined as follows: In Section~\ref{sec:model}, we describe the theory behind the \ahod\ framework. In Section~\ref{sec:simulation}, we briefly describe \textsc{AbacusSummit} simulations that the \ahod\ is currently built on. In Section~\ref{sec:algorithm}, we present the core algorithm and its many optimizations. In Section~\ref{sec:application}, we showcase the first example application of \ahod, specifically to model the full-shape redshift-space clustering of CMASS galaxies. \red{In Section~\ref{sec:eboss}, we showcase the second example application, where we model clustering of the multi-tracer eBOSS sample. In Section~\ref{sec:discuss}, we discuss some interesting aspects of our analyses, especially with regard to the ``lensing is low'' issue, and compare to previous analyses.} Finally, we conclude in Section~\ref{sec:conclude}.

\section{The extended HOD framework}
\label{sec:model}
In this section, we introduce the extended multi-tracer HOD framework, starting with the baseline HOD model for the three dark-time tracers expected for DESI: LRG, ELG, and QSO. Then we describe extensions to the baseline model, including satellite profile variations, velocity bias, assembly bias, and environment-based secondary bias. 

\subsection{The baseline HOD model}
The baseline HOD for LRGs comes from the 5-parameter model described in \citet{2007bZheng}, which gives the mean expected number of central and satellite galaxies per halo given halo mass:
\begin{align}
    \bar{n}_{\mathrm{cent}}^{\mathrm{LRG}}(M) & = \frac{1}{2}\mathrm{erfc} \left[\frac{\log_{10}(M_{\mathrm{cut}}/M)}{\sqrt{2}\sigma}\right], \label{equ:zheng_hod_cent}\\
    \bar{n}_{\mathrm{sat}}^{\mathrm{LRG}}(M) & = \left[\frac{M-\kappa M_{\mathrm{cut}}}{M_1}\right]^{\alpha}\bar{n}_{\mathrm{cent}}^{\mathrm{LRG}}(M),
    \label{equ:zheng_hod_sat}
\end{align}
where the five parameters characterizing the model are $M_{\mathrm{cut}}, M_1, \sigma, \alpha, \kappa$. $M_{\mathrm{cut}}$ characterizes the minimum halo mass to host a central galaxy. $M_1$ characterizes the typical halo mass that hosts one satellite galaxy. $\sigma$ describes the steepness of the transition from 0 to 1 in the number of central galaxies. $\alpha$ is the power law index on the number of satellite galaxies. $\kappa M_\mathrm{cut}$ gives the minimum halo mass to host a satellite galaxy.
We have added a modulation term $\bar{n}_{\mathrm{cent}}^{\mathrm{LRG}}(M)$ to the satellite occupation function to remove satellites from halos without centrals. 

In the baseline implementation, the actual number of central galaxy per halo is drawn from a Bernoulli distribution with mean equal to $\bar{n}^{\mathrm{LRG}}_{\mathrm{cent}}(M)$, and the actual number of satellite galaxies is drawn from a Poisson distribution with mean equal to $\bar{n}^{\mathrm{LRG}}_{\mathrm{sat}}(M)$. The central galaxy is assigned to the center 
of mass of the largest sub-halo, with the velocity vector also set to that of the center 
of mass of the largest sub-halo. Satellite galaxies are assigned to particles of the 
halo with equal weights. 

Similarly, the baseline HOD for ELGs is based on the parametric model described in \citet{2020Avila} and \citet{2020Alam}:
\begin{align}
    \bar{n}_{\mathrm{cent}}^{\mathrm{ELG}}(M)  &=  2 A \phi(M) \Phi(\gamma M)  + & \nonumber \\  
    \frac{1}{2Q} & \left[1+\mathrm{erf}\left(\frac{\log_{10}{M_h}-\log_{10}{M_{\mathrm{cut}}}}{0.01}\right) \right],  \label{eq:NHMQ}
\end{align}
where
\begin{align}
\phi(x) &=\mathcal{N}(\log_{10}{ M_{\mathrm{cut}}},\sigma_M), \label{eq:NHMQ-phi}\\
\Phi(x) &= \int_{-\infty}^x \phi(t) \, dt = \frac{1}{2} \left[ 1+\mathrm{erf} \left(\frac{x}{\sqrt{2}} \right) \right], \label{eq:NHMQ-Phi}\\
A &=p_{\rm max}  -1/Q.
\label{eq:alam_hod_elg}
\end{align}
whereas the satellite occupation continues to adopt the power law form of Equation~\ref{equ:zheng_hod_sat}. Compared to Equ~9-12 in \citet{2020Alam}, we modified the definition of $A$ by skipping the denominator for ease of computation. We do not notice any significant change to the functional form of the central occupation. This baseline HOD form for ELGs is also confirmed in simulation and semi-analytic model approaches by studies such as \citet{2021eHadzhiyska} and \citet{2020Gonzalez-Perez}. 

The baseline QSO HOD adopts the same functional form as the LRGs, referring back to Equation~\ref{equ:zheng_hod_cent}-\ref{equ:zheng_hod_sat}. We also implement a overall amplitude parameter for each tracer to account for incompleteness. 

In the current version of \ahod, we enforce that each halo can only host at most one central galaxy, which can be of any tracer type. Similarly, each particle
can also host at most one satellite galaxy, which can be of any tracer type. We do not enforce any type of central-satellite conformity or 2-halo conformity. 

In the following subsections, we introduce our framework of extending the baseline HODs with physical generalizations. 

\subsection{The satellite profile generalizations}

In the baseline implementation, we assume that the 1-halo distribution of satellites tracks the halo density profile, where we assign satellites to halo particles with equal weight. \citet{2019Bose} used hydrodynamical simulations to show that this is a reasonable assumption for mass-selected galaxy samples. In this section, we relax that assumption by introducing several physically motivated generalizations that allow satellite profile to deviate from that of the dark matter halo. Our generalizations are based on re-weighting existing particles in the halo instead of simply moving galaxies, thus preserving Newtonian physics. These generalizations correspond to parameters $s$, $s_p$, and $s_v$, which were previously introduced in Section~3.2 of \citet{2018Yuan} and Section~2.2 of \citet{2021Yuan}. We summarize the key ideas here. 

For instance, the $s\in [-1, 1]$ parameter biases the satellite profile radially, where $s>0$ corresponds to the distribution of satellites being puffier than the halo profile, and $s<0$ corresponding to satellite distribution being more concentrated. The $s$ parameter works by first ranking all particles within a halo by their radial distance to halo center, and then assigning to each particle a weight that linearly depends on rank. Specifically, the probability for the $i$-th ranked particle to host a satellite galaxy is given by
\begin{equation}
p_i = \frac{\bar{n}_\mathrm{sat}}{N_p}\left[1+s(1 - \frac{2r_i}{N_p-1})\right],\ \ \ \ (i = 0, 1, 2, ..., N_p - 1)
\label{equ:pi_s}
\end{equation}
where $N_p$ is the number of particles in the halo, and $r_i$ is the rank of the $i$-th particle. 

Similarly, we introduce the $s_v$ parameter, which biases the satellite profile based on particle peculiar velocity, and the $s_p$ parameter, which is based on particle perihelion distance. A detailed description of these two parameters, including how to estimate perihelion distance, can be found in Section~3.3 and 3.4 of \citet{2018Yuan}. There are several motivations in including these satellite profile generalization parameters. Baryonic processes can bias the concentration of baryons within the dark matter potential well \citep[e.g.][]{2010Duffy, 2010Abadi, 2017Chua, 2017Peirani, 2020Amodeo}. Splashback and infall can introduce biases for satellites with eccentric orbits \citep{2014Behroozi, 2014Diemer, 2014Adhikari, 2015More}. In our previous analyses, we have also used the $s_v$ parameter as an alternative but more physical model for satellite velocity bias. We discuss this approach further in the following subsection. 

\subsection{Velocity bias}

While velocity measurements are not relevant for the study of galaxy positions in real space, the velocities do become entangled with the line-of-sight (LOS) positions in redshift-space due to redshift-space distortions \citep{1987Kaiser}. Thus, to model the redshift-space clustering of galaxies with high fidelity, we need to introduce a more flexible model of galaxy velocities.  

In the baseline implementation, we assume the velocity of the central galaxy to be the bulk velocity of the largest subhalo, and the velocity of the satellite galaxies to be the same as their host particles. Following observational evidence presented in \citet{2014Reid} and \citet{2015aGuo}, we introduce a velocity bias model that allows for deviations in central and satellite velocities from the underlying dark matter. First we add an additional Gaussian scatter to the LOS component of the central galaxy velocity, with width equal to the halo particle velocity dispersion. The central galaxy velocity along the LOS is thus given by 
\begin{equation}
    v_\mathrm{cent, z} = v_\mathrm{L2, z} + \alpha_c \delta v(\sigma_{\mathrm{LOS}}),
    \label{equ:alphac}
\end{equation}
where $v_\mathrm{L2, z}$ denotes the line-of-sight component of the subhalo velocity, $\delta v(\sigma_{\mathrm{LOS}})$ denotes the Gaussian scatter, and $\alpha_c$ is the central velocity bias parameter, which modulates the amplitude of the central velocity bias effect. By definition, $\alpha_c$ is non-negative, and $\alpha_c = 0$ corresponds to no central velocity bias.

For the satellite galaxies, we allow for their peculiar velocities to be systematically higher or lower than their host particle velocities. This satellite velocity bias effect is given by 
\begin{equation}
    v_\mathrm{sat, z} = v_\mathrm{L2, z} + \alpha_s (v_\mathrm{p, z} - v_\mathrm{L2, z}),
    \label{equ:alpha_s}
\end{equation}
where $v_\mathrm{p, z}$ denotes the line-of-sight component of particle velocity, and $\alpha_s$ is the satellite velocity bias parameter. $\alpha_s = 1$ corresponds to no satellite velocity bias. 

While this $(\alpha_c, \alpha_s)$ model presented here is the most common implementation of velocity bias, it does break Newtonian physics by modifying satellite velocity without modifying its position. In \citet{2021Yuan}, we used a different implementation of velocity bias, where we replaced the $\alpha_s$ parameter with the $s_v$ parameter. The $s_v$ parameter simultaneously modulates the radial distribution and peculiar velocity of the satellite by preferentially assigning satellites to particles with higher or lower peculiar velocities, thus ensuring Newtonian orbits. 
However, the observed velocity bias is not necessarily exclusively due to halo physics, but could also be due to redshift systematics, which decouples the satellite velocities from satellite position. In this analysis, we use the more common $(\alpha_c, \alpha_s)$ prescription to marginalize over such observation systematics. 

\subsection{Secondary biases}

Following lessons learned in \citet{2021Yuan}, we extend the standard HOD with two secondary dependencies, one on halo concentration (assembly bias), and one on the local environment. The concentration dependency describes the classical galaxy assembly bias effect, where the HOD model depends on the assembly history of the halo (encoded by halo concentration) in addition to halo mass. The local environment dependency is more novel, but it was found to be an necessary tracer of secondary biases in both simulations \citep{2020Hadzhiyska, 2020Xu} and observations \citep{2021Yuan}.

There are multiple frameworks for introducing these secondary dependencies. \citet{2016Hearin} and \citet{2021Yuan} both adopt a ``galaxy swap" approach, where galaxies are swapped between halos of different secondary properties. This approach naturally conserves the total number density of galaxies, but it tends to be computationally expensive. In the \ahod\ package, we adopt a different but computationally cheaper approach, where we analytically mix the secondary property with halo mass. This approach was used previously both in \citet{2019Walsh} and \citet{2020Xu}. We specifically follow the analytic prescription of \citet{2020Xu}, 
where the secondary halo property is directly tied to the two mass parameters in the baseline HOD, $M_{\mathrm{cut}}$ and $M_1$:
\begin{align}
 \log_{10} M_{\mathrm{cut}}^{\mathrm{mod}} & = \log_{10} M_{\mathrm{cut}} + A_\mathrm{cent}(c^{\mathrm{rank}} - 0.5) + B_\mathrm{cent}(\delta^{\mathrm{rank}} - 0.5) \\
 \log_{10} M_{1}^{\mathrm{mod}} & = \log_{10} M_{1} + A_\mathrm{sat}(c^{\mathrm{rank}} - 0.5) + B_\mathrm{sat}(\delta^{\mathrm{rank}} - 0.5)
 \label{equ:AB}
\end{align}
where $c$ and $\delta$ are the halo concentration and local 
overdensity, respectively. These secondary properties are ranked within narrow
halo mass bins, and the resulting ranks $c^{\mathrm{rank}}$ and $\delta^{\mathrm{rank}}$ are normalized to range from 0 to 1, with 0.5 corresponding to the median. For example, $c^{\mathrm{rank}} > 0.5$ corresponds to a halo with above median concentration, $c^{\mathrm{rank}} < 0.5$ corresponds to a halo with below median concentration, and the same logic applies for the environment rank $\delta^{\mathrm{rank}}$. The tetrad  $(A_\mathrm{cent}, B_\mathrm{cent}, A_\mathrm{sat}, B_\mathrm{sat})$ form the four parameters describing the amplitude of secondary biases in our HOD model. No secondary bias corresponds to all four parameters being equal to zero. 
We should also point out that in this prescription, the sign of the secondary bias parameters goes in the opposite direction of the secondary bias parameters in \citet{2021Yuan}. For example, in this new prescription, a positive $A_\mathrm{cent}$ would increase $\log M_\mathrm{cut}$ for the more concentrated ($c^{\mathrm{rank}} > 0.5$) halos, which reduces the number of galaxies in more concentrated halo at fixed halo mass (refer to Equation~\ref{equ:zheng_hod_cent} and Equation~\ref{eq:NHMQ}). Whereas in the model used in \citet{2021Yuan}, a positive $A$ increases the number of galaxies in more concentrated halos. The same logic applies to the environment-based bias, where we switch $A$ with $B$. So to summarize so far, positive $A$ ($B$) means high-concentration (environment) halos have fewer galaxies and low-concentration (environment) halos have more galaxies at fixed mass.

The concentration $c$ is defined as the ratio $c = r_{90}/r_{25}$, where $r_x$ refers to the radius that encloses $x\%$ of the total halo mass. 
The local overdensity $\delta$ is calculated in a very similar fashion to \citet{2021Yuan}. First, for each halo, we compute the total enclosed mass of the neighboring halos, where a neighbor halo is defined to be within 5$h^{-1}$Mpc but beyond the halo radius $r_{98}$. Then we divide the enclosed mass by the average enclosed mass to obtain the local overdensity. Mathematically, we express this definition as 
\begin{equation}
    \delta = \frac{M(r_{98} < r < 5h^{-1}\mathrm{Mpc})}{\langle M(r_{98} < r < 5h^{-1}\mathrm{Mpc})\rangle} - 1,
\end{equation}
where $M$ denotes the enclosed mass of the neighboring halos. 

\begin{figure}
    \centering
    \begin{subfigure}[]{0.4\textwidth}
    \hspace*{-0.8cm}
    \includegraphics[width = 3.4in]{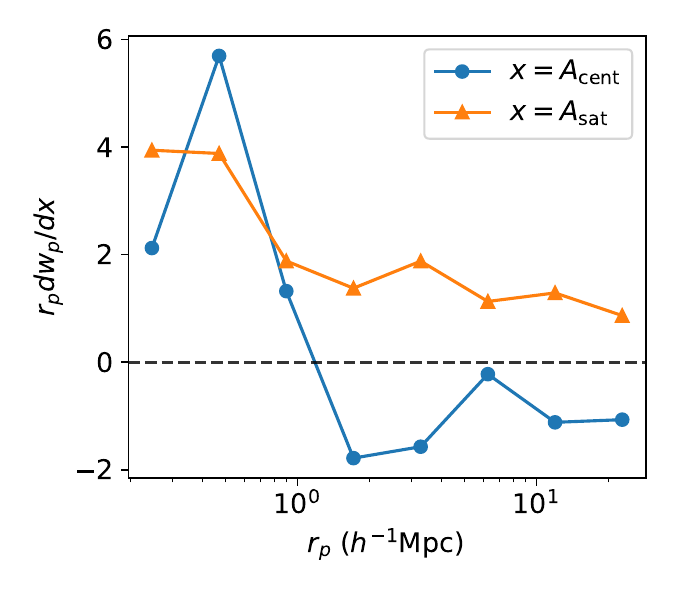}
     \vspace{-0.8cm}
    \caption{Derivative against assembly bias parameters $A_\mathrm{cent}$ and $A_\mathrm{sat}$}
    \label{fig:derivA}
    \end{subfigure}
    \begin{subfigure}[]{0.4\textwidth}
    \hspace*{-1.1cm}
    \includegraphics[width = 3.5in]{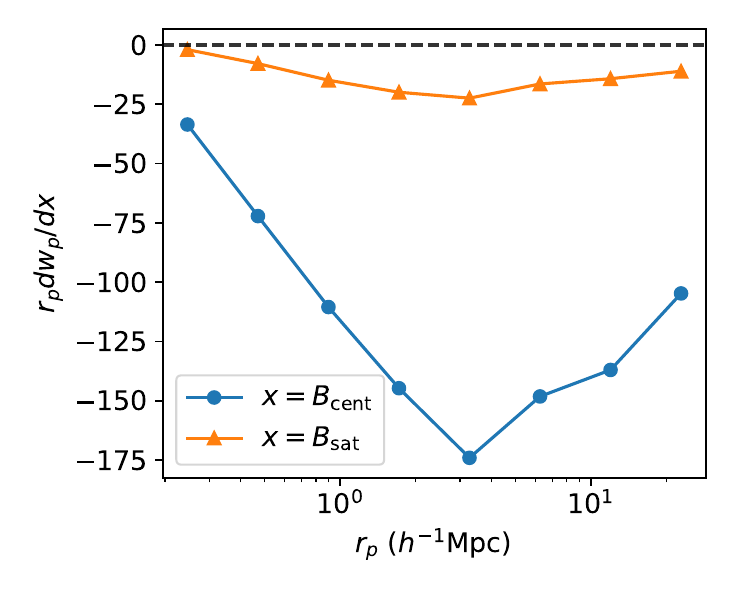}
     \vspace{-0.8cm}
    \caption{Derivative against environment-based bias parameters $B_\mathrm{cent}$ and $B_\mathrm{sat}$}
    \label{fig:derivB}
    \end{subfigure}
    \caption{Derivatives of the projected galaxy correlation function $w_p$ (Equation~\ref{equ:wp_def}) against the secondary bias parameters. This is to show help the readers gain intuition on how the four secondary bias parameters impact the predicted galaxy clustering. The top panel shows the derivative against the assembly bias parameters $A_\mathrm{cent}$ and $A_\mathrm{sat}$, whereas the bottom panel shows the derivatives against the environment-based secondary bias parameters $B_\mathrm{cent}$ and $B_\mathrm{sat}$. }
    \label{fig:derivs}
\end{figure}

To gain intuition on how the secondary biases impact galaxy clustering, we show the derivatives of the galaxy projected clustering function $w_p$ (see Equation~\ref{equ:wp_def}) against each of the four secondary bias parameters in Figure~\ref{fig:derivs}. The top panel shows the derivatives against assembly bias parameters, whereas the bottom panel shows the derivatives against the environment-based secondary bias parameters. For the assembly bias parameters, we see that they are more important at very small scales $r_p < 1h^{-1}$Mpc while their effect diminishes at larger scales. At $r_p < 1h^{-1}$Mpc, the clustering is dominated by the 1-halo term, i.e. central-satellite clustering and satellite-satellite clustering. It makes sense that both the central and satellite derivatives are positive in this regime. Specifically, for a positive assembly bias parameter, the more concentrated halos ($c_{\mathrm{rank}} > 0.5$) correspond to a higher $M_{\mathrm{cut}}^{\mathrm{mod}}$ and $M_{1}^{\mathrm{mod}}$, and higher $M_{\mathrm{cut}}^{\mathrm{mod}}$ and $M_{1}^{\mathrm{mod}}$ mean fewer centrals and satellites for those halos. Thus, positive assembly bias parameters disfavor more concentrated halos and favor less concentrated halos. By putting more centrals and satellite into less concentrated halos, positive assembly bias boosts the central-satellite and satellite-satellite pair counts at the halo-size scale, which is typically $0.1-1h^{-1}$Mpc, thus boosting the clustering at those scales.

The bottom panel of Figure~\ref{fig:derivs} shows that both environment-based bias parameters have negative derivatives in the projected galaxy clustering. This makes sense because for positive environment-based bias parameters, the $M_{\mathrm{cut}}^{\mathrm{mod}}$ and $M_{1}^{\mathrm{mod}}$ parameters for halos in denser environments are increased. Thus, positive environment bias parameters favor halos in less dense environments to host galaxies, which leads to lower galaxy pair counts, thus the lower clustering amplitude. We see that the effect is strongest at around $r_p\sim 3h^{-1}$Mpc, which is the characteristic scale that our environment is defined at. Compared to the concentration-based assembly bias parameters, it is clear that while assembly bias mostly impacts clustering in the 1-halo term, environment-based secondary bias affects mostly the 2-halo term and extends out to much larger scales. 

The clustering signature of these secondary biases is ultimately the combined effect of occupational biases such as the ones modeled in this section, and of how halo clustering depends on secondary properties, an effect known as halo assembly bias \citep[e.g.][]{2005Gao, 2007Croton}. Specifically, at fixed halo mass, high-environment and high-concentration halos tend to be more clustered, so when one varies the galaxy occupation as a function of these secondary parameters, one also changes the galaxy clustering by favoring more or less clustered halos. We refer the readers to \citet{2021bHadzhiyska} for a detailed presentation of the interaction between galaxy occupational variation and halo assembly bias. 

\section{The \textsc{AbacusSummit} simulations}
\label{sec:simulation}

In principle, our extended model is not simulation specific, as long as the simulation outputs a halo and particle catalog. Currently, the \ahod\ code package is specifically set up for the \textsc{AbacusSummit} simulation suite, which is a set of large, high-accuracy cosmological N-body simulations using the \textsc{Abacus} N-body code \citep{2019Garrison, 2021bGarrison}, designed to meet the Cosmological Simulation Requirements of the Dark Energy Spectroscopic Instrument (DESI) survey \citep{2013arXiv1308.0847L}. \textsc{AbacusSummit} consists of over 150 simulations, containing approximately 60 trillion particles at 97 different cosmologies. A typical base simulation box contains $6912^3$ particles within a $(2h^{-1}$Gpc$)^3$ volume, which yields a particle mass of $2.1 \times 10^9 h^{-1}M_\odot$. \footnote{For more details, see \url{https://abacussummit.readthedocs.io/en/latest/abacussummit.html}}

The set of example fits presented in this paper are done primarily using the $z = 0.5$ slice of the \verb+AbacusSummit_base_c000_ph000+
box, which is $(2h^{-1}$Gpc$)^3$ in volume and adopts the Planck 2018 $\Lambda$CDM cosmology ($\Omega_c h^2 = 0.1200$, $\Omega_b h^2 = 0.02237$, $\sigma_8 = 0.811355$, $n_s = 0.9649$, $h = 0.6736$, $w_0 = -1$, and $w_a = 0$). 

The {\sc CompaSO} halo finder is a highly efficient on-the-fly group finder specifically designed
for the \textsc{AbacusSummit} simulations \citep{2021Hadzhiyska}. 
{\sc CompaSO} builds on the existing 
spherical overdensity (SO) algorithm
by taking into consideration the tidal radius
around a smaller halo before competitively
assigning halo membership to the particles
in an effort to more effectively deblend halos.
Among other features, the {\sc CompaSO} finder also
allows for the formation of new halos on the 
outskirts of growing halos, which alleviates
a known issue of configuration-space halo 
finders of failing to identify halos close to
the centers of larger halos. 

We also run a post-processing ``cleaning'' procedure that leverages the halo merger trees to ``re-merge'' a subset of halos. This is done both to remove over-deblended halos in the spherical overdensity finder, and to intentionally merge physically associated halos that have merged and then physically separated. An example of such dissociation is what is known as splashback \citep[e.g.][]{2014Diemer, 2015bMore, 2016More}, where halos that were once part of a larger halos have since exited following at least one orbital passage within their former hosts. In \citet{2021Bose}, we find that remerging such halos signicantly improves the fidelity of the halo catalog, and the resulting ``cleaned'' halo catalog achieves significantly better fits on data in an HOD analysis. The fits presented in later sections of this paper are carried out with the cleaned halo catalogs. 

\section{\textsc{AbacusHOD}: core algorithm and optimizations}
\label{sec:algorithm}
The \ahod\ module loads the halo and particle catalogs from the \textsc{AbacusSummit} 
simulations and outputs multi-tracer mock galaxy catalogs. This code is designed particularly for efficient HOD parameter searches, in which
many HOD parameter sets will be requested in quick succession. In this section, we describe the core algorithm and some key optimizations implemented to maximize efficiency. 

The mock galaxy generation is divided into two stages, a preparation stage and an HOD stage. The preparation stage needs to be run first and serves to process the raw halo and particle files, front-loading all the expensive I/O and optimizes the simulation data for the second much faster HOD evaluation. 

\subsection{The preparation stage}
One key objective of the preparation stage is to downsample the halos and particles from the simulation box. This is because the speed of evaluating an HOD scales roughly linearly with the number of halos and particles passed to the HOD code, barring a small amount of overheads. Thus, by optimally downsampling the halos and particles in the preparation stage, we can substantially increase the efficiency of each evaluation of the HOD stage. 

To this end, we implement a mass-dependent downsampling of the halos and particles. Specifically, we use a sigmoid function to aggressively downsample low mass halos while preserving most halos at high mass, where the turn off mass and the turn off rate depends on the tracer type. 
For the particles, we apply a uniform downsampling for all masses in addition to the sigmoid turn off. The goal is to reduce the number of particles until the number of particles per halo is only 10-100 times the expected number of satellites. 
By default, we implement two sets of downsampling filters, one designed for CMASS LRGs, and the other designed for ELGs and QSOs. The second filter goes to substantially lower halo mass and thus contains a significantly larger number of halos, resulting in lower performance in the HOD stage. The user should use these downsampling filters as a point of reference and customize the downsampling function as needed. The default filter for LRGs is shown in Equation~\ref{equ:downsampling}. 

Another key objective of the preparation stage is to precompute all the halo and particle properties necessary for the HOD model and concatenate them into a large contiguous array, Along with relevant halo and particle properties, the code also marks each halo and particle with a random number. The random numbers are for drawing from the central Bernoulli distribution and the satellite Poisson distribution. Pre-generating random numbers for all halos and particles not only reduces the computational cost when running HOD chains, but also carries the additional benefits of removing realization noise and making the mocks reproducible. The removal of realization noise also allows for calculation of more accurate derivatives of summary statistics against HOD parameters. 

\subsection{The HOD stage}
The HOD stage centers around the \textsc{AbacusHOD} class object, which loads the downsampled halo and particle catalogs from the preparation stage onto memory when initialized, and then takes input HOD parameters and returns galaxy mock catalogs. In each HOD call, the centrals and satellites are generated separately and then concatenated into one unified output dictionary. 

To efficiently generate mocks given an input HOD, we adopt a multi-threaded two-pass memory-in-place algorithm accelerated with \textsc{numba}. To maximize the efficiency of multi-threading, the first pass serves to exactly determine the galaxy generating workload, and evenly partition the workload across all threads and pre-allocate the amount of memory needed for each thread. It does so by looping through the halos (particles) and calculating the number of centrals (satellites) to be generated for each halo by comparing the mean number of centrals (satellites) of that halo to its corresponding pre-generated random numbers. Then, it allocates an empty array for all galaxies to be generated, including their properties (position, velocity, and etc.). The galaxy array is then evenly partitioned by the number of threads and each partition is assigned to a thread. Finally on the second pass, each thread loops through its assigned halos and particles and fills out the galaxy array. 
A two-pass approach achieves significantly better performance than a typical brute-force approach by storing halos/particles data in memory, avoiding the costly operation of copy-pasting the entire galaxy array every time new galaxies need to be added. 

We further accelerate the halo and particle for-loops with \textsc{numba} no-python compiler to compile the slower Python code into faster machine code, and we take advantage of all available processor cores with by multi-threading over the halo and particle loops. 

\subsection{Utility functions}
As part of the \ahod\ package, we also provide additional utility functions that are commonly needed for HOD analyses. These include several 2-point correlation function (2PCF) calculators, a galaxy-galaxy lensing calculator, and sampling scripts. 

The provided 2PCF calculators are based on the highly-optimized grid-based \textsc{Corrfunc} code \citep{2020Sinha}. We further optimize the performance of \textsc{Corrfunc} to match that of the HOD code. 
The galaxy-galaxy lensing calculator is highly optimized compared to the \textsc{halotools} lensing calculator \citep{2017Hearin}. It takes advantage of the fact that the g-g lensing measurement $\Delta \Sigma$ is a linearly combination of the $\Delta\Sigma$ at each galaxy position. It pre-computes and saves the $\Delta\Sigma$ at each halo and particle position. For each HOD evaluation, it simply conducts a weighted sum of the halos and particles with galaxy weights given by the HOD. This lensing calculator is suited for fitting lensing measurements, not for making a single lensing prediction due to the high cost of pre-computing halo and particle $\Delta\Sigma$. 

We provide two popular methods of HOD sampling with \ahod. The first one is an MCMC based script with \textsc{emcee} \citep{2013Foreman}, and the second one is a nested sampling script with \textsc{dynesty} \citep{2018Speagle}. We recommend the nested sampling script for much higher sampling efficiency and the natural calculation of model evidence, which is an essential metric for model comparisons. These scripts can be found at \texttt{abacusutils/scripts/hod/}. We provide a case study using the nested sampler in later sections. 

\subsection{Performance}
A key characteristic of the \ahod\ code is its high efficiency. In this sub-section, we offer a performance benchmark of the code. Our test system consists of two Intel Xeon Gold 5218 CPUs clocked at 2.30GHz, for a total of 32 cores a single node, and 256GB of DDR4 RAM. We use \textsc{Python} version 3.7.9, \textsc{Numpy} version 1.19.2, and \textsc{numba} version 0.51.2. 
For the test runs, we process a single \textsc{AbacusSummit} base simulation box at Planck cosmology, specifically the \texttt{AbacusSummit\_base\_c000\_ph000} box. We load the cleaned \textsc{CompaSO} halos at the $z = 0.5$ snapshot and downsample the halos and particles using the default filters provided in the package, as shown in Equation~\ref{equ:downsampling}:
\begin{align}
    f_\mathrm{halos} & = \frac{1}{1+0.1\exp(-(\log M_h - 13.3)\times 25)}, \nonumber\\
    f_\mathrm{particles} & = \frac{4}{200+\exp(-(\log M_h - 13.7)\times 8)}.
    \label{equ:downsampling}
\end{align}
The downsampling reduces the total number of halos from $3.99\times 10^8$ down to $6.18\times 10^6$, and the total number of subsample particles from $3.15\times 10^8$ down to $1.77\times 10^7$. We note that, given the small number of satellites produced, it is likely possible to further downsample the particles. However, we do not further optimize the particle sample for this analysis since generating satellites is not the performance bottleneck in our tests. 

We pick a fiducial baseline HOD prescription $\log_{10} M_\mathrm{cut} = 12.8$, $\log_{10}M_1 = 14.0$, $\sigma = 0.5$, $\alpha = 1.0$, and $\kappa = 0.5$, roughly resembling a CMASS-like sample. The construction of the \ahod\ object, i.e. loading of halo and particle subsamples onto memory, takes approximately $10$ seconds. Then, we run the HOD code once to compile the code with \textsc{numba}, which takes around $10$ seconds. Then we repeat each HOD run 20 times, and take the average run time. Here we showcase how the run time of the mock generation and the 2PCF calculator depends on the number of threads, and the number density of galaxies. Specifically for the 2PCF calculator, we compute $\xi(r_p, \pi)$ with 8 logarithmic bins in $r_p$ between $0.169h^{-1}$Mpc and 30$h^{-1}$Mpc, and 6 linear bins in $\pi$ between 0 and 30$h^{-1}$Mpc. 

\begin{figure}
    \centering
    \hspace*{-0.6cm}
    \includegraphics[width = 3.2in]{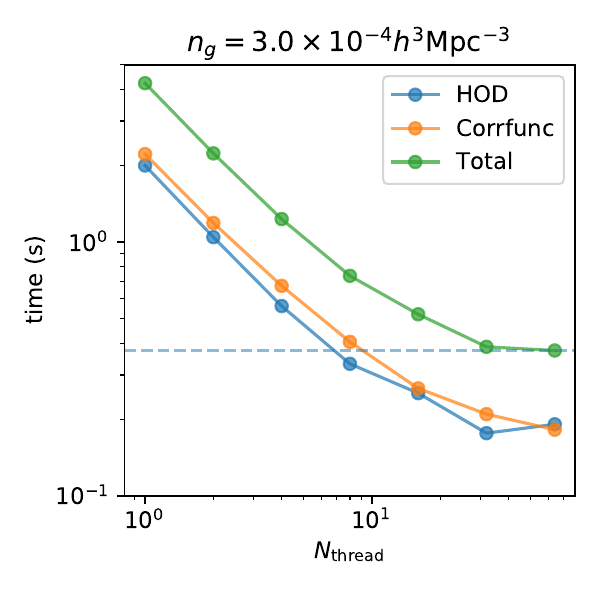}
    \vspace{-0.3cm}
    \caption{Timing of the HOD evaluation and the 2PCF calculator as a function of the number of threads when running on a 32 core node, at fixed galaxy number density (CMASS average density). Both calculations are scalable. The dashed line shows the minimum total timing, at just below 0.4 seconds per HOD call. The timing plateaus above 32 threads, where running multiple threads per core (hyper-threading) provides little to no performance gain.}
    \label{fig:timing_thread}
\end{figure}

Figure~\ref{fig:timing_thread} shows how the timing of an HOD evaluation and the 2PCF calculation scales with the number of threads, where the galaxy number density is fixed at BOSS CMASS average density, $3\times 10^{-4}h^3$Mpc$^{-3}$. We see that both calculations are highly scalable at below $N_\mathrm{thread} < 32$, above which we start to lose per thread efficiency due to hyper-threading, i.e. running more than one thread per core provides little to no gain. The best-case timing for the HOD code (mock generation) is 0.17 seconds with 32 threads. For the 2PCF calculator, the best timing is 0.18 seconds with 64 threads.
\begin{figure}
    \centering
    \hspace*{-0.6cm}
    \includegraphics[width = 3.2in]{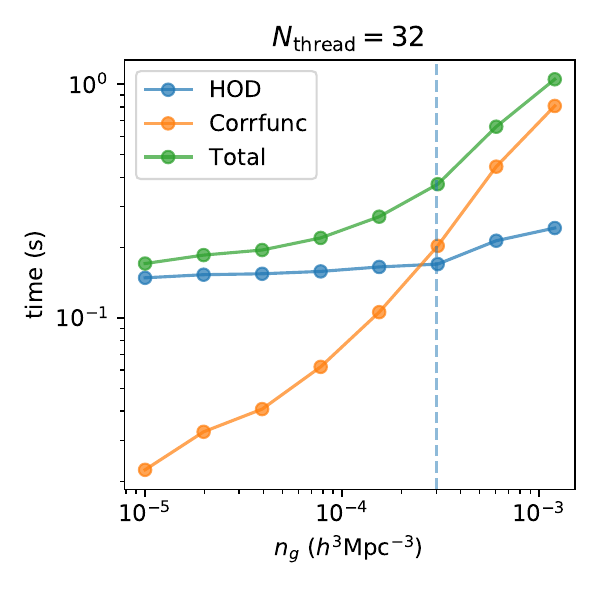}
    \vspace{-0.3cm}
    \caption{Timing of the HOD code and the 2PCF calculator as a function of galaxy number density at $N_\mathrm{thread} = 32$. The timing of the HOD code is largely independent of number density, whereas the 2PCF timing scales roughly linearly with number density. }
    \label{fig:timing_density}
\end{figure}
Figure~\ref{fig:timing_density} shows how the timing scales with the galaxy number density, at fixed number of threads ($N_\mathrm{thread} = 32$). The dashed line denotes the CMASS average density. We see that the HOD code is largely independent of the galaxy number density, since the HOD code timing depends on the number of halos and particles considered, not the number of galaxies produced by that sweep.
 The 2PCF code scales roughly linearly with the number density. This suggests that the 2PCF timing is likely dominated by linear overheads, such as gridding. We have optimized the 2PCF calculator by choosing the optimal grid size. The user may find further tuning of the grid size necessary depending on the simulation size and sample density. 

The performance of \ahod\ in a real world application is highly dependent on the hardware, simulations, and the summary statistics. Our tests are conducted on a single node system with generous memory and relatively fast processors. On a cluster system like cori at NERSC, the user might also benefit from chain-level parallelization instead of focusing on the timing of a single HOD evaluation. We also did not present the timing for lower mass tracers such as ELGs, but internal testings show that evaluating an eBOSS-like ELG HOD is approximately 2-3 times slower than evaluating a CMASS-like LRG HOD. 
An important limiting factor in the performance is the summary statistics calculator. While we provide fast calculators for 2PCF and galaxy-galaxy lensing, anything beyond these statistics remains the responsibility of the user for now. 

Compared to other existing HOD implementations, \ahod\ is $\sim 100$ times faster than other particle-based algorithms, including the \textsc{GRAND-HOD} code we developed in \citet{2018Yuan} and the \textsc{halotools} implementation in \citet{2013Behroozi}. The tabulated HOD (e.g. \textsc{TabCorr}\footnote{\url{https://github.com/johannesulf/TabCorr}}) approach can be of similar or better performance by pre-computing all the halo clustering and then convolving halo occupation with the pre-computing clustering \citep{2016Zheng}. The tabulated HOD approach achieves good performance by sacrificing model flexibility, where extending a tabulated HOD models with additional features could add significant complexities to the pre-tabulation and the computation of the convolution. 

\section{Application to BOSS LRG clustering}
\label{sec:application}
In this section, we apply the \ahod\ package to fitting the BOSS CMASS and LOWZ LRG clustering. Besides serving as an example application for the \ahod\ package, we also compare the constraining power of the projected clustering and the redshift-space clustering. \red{We test the necessity of various HOD extensions for this galaxy sample, which leads to implications in the galaxy-galaxy lensing tension \citep{2017Leauthaud} (discussed in section~\ref{subsec:lensing}).} For brevity, we lead with a detailed analysis on CMASS but only summarize the key results of the LOWZ analysis. 

\subsection{BOSS galaxy sample}
\label{sbsec:cmass}
The Baryon Oscillation Spectroscopic Survey \citep[BOSS; ][]{2012Bolton, 2013Dawson} is part of the SDSS-III programme \citep{2011Eisenstein}. BOSS Data Release 12 (DR12) provides redshifts for 1.5 million galaxies in an effective area of 9329 square degrees divided into two samples: LOWZ and CMASS. The LOWZ galaxies are selected to be the brightest and reddest of the low-redshift galaxy population at $z < 0.4$, whereas the CMASS sample is designed to isolate galaxies of approximately constant mass at higher redshift ($z > 0.4$), most of them being also Luminous Red Galaxies \citep[LRGs,][]{2016Reid, 2016Torres}. The survey footprint is divided into chunks which are covered in overlapping plates of radius $\sim 1.49$ degrees. Each plate can house up to 1000 fibres, but due to the finite size of the fibre housing, no two fibres can be placed closer than $62$ arcsec, referred to as the fibre collision scale \citep{2012Guo}. 

For the CMASS analysis, we limit our measurements to the galaxy sample between redshift $0.46 < z < 0.6$ in DR12, whereas for LOWZ, we adopt redshift range $0.15 < z < 0.4$. We choose these moderate redshift ranges for completeness and to minimize the systematics due to redshift evolution. Applying this redshift range to both the north and south galactic caps gives a total of approximately 600,000 galaxies in our CMASS sample, and just under 400,000 galaxies in our LOWZ sample. The average galaxy number density is given by $n_\mathrm{CMASS} = (3.01\pm 0.03)\times 10^{-4} h^{3}$Mpc$^{-3}$ for CMASS and $n_\mathrm{LOWZ} = (3.26\pm 0.03)\times 10^{-4} h^{3}$Mpc$^{-3}$ for LOWZ. 

We consider two key 2-point statistics on the data. The first is the redshift-space 2PCF $\xi(r_p, \pi)$, which can be computed using the \citet{1993Landy} estimator:
\begin{equation}
    \xi(r_p, \pi) = \frac{DD - 2DR + RR}{RR},
    \label{equ:xi_def}
\end{equation}
where $DD$, $DR$, and $RR$ are the normalized numbers of data-data, data-random, and random-random pair counts in each bin of $(r_p, \pi)$, and $r_p$ and $\pi$ are transverse and line-of-sight (LOS) separations in comoving units. For this paper, we choose a coarse binning to ensure reasonable accuracy on the covariance matrix, with 8 logarithmically-spaced bins between 0.169$h^{-1}$Mpc and 30$h^{-1}$Mpc in the transverse direction, and 6 linearly-spaced bins between 0 and 30$h^{-1}$Mpc bins along the line-of-sight direction. The same binning is used for both the CMASS and LOWZ samples. 

The second statistic is the projected galaxy 2PCF, commonly referred to as $w_p$. It is simply defined as the line-of-sight integral of the redshift-space $\xi(r_p, \pi)$,
\begin{equation}
w_p(r_p) = 2\int_0^{\pi_{\mathrm{max}}} \xi(r_p, \pi)d\pi,
\label{equ:wp_def}
\end{equation}
where $\pi_{\mathrm{max}} = 30 h^{-1}$Mpc. We use a finer binning for $w_p$, with a total of 18 bins between 0.169$h^{-1}$Mpc and 30$h^{-1}$Mpc. 

We have corrected the data for fibre collision effects following the method of \cite{2012Guo}, by separating galaxies into collided and decollided populations and assuming those collided galaxies with measured redshifts in the plate-overlap regions are representative of the overall collided population. The final corrected correlation function can be obtained by summing up the contributions from the two populations. However, scales below 0.5$h^{-1}$Mpc likely still suffer from systematics even after the correction, and they show a turn off that is qualitatively inconsistent with theoretical expectations and simulations. Thus, we remove the first three bins in $w_p$, and the first column of $\xi(r_p, \pi)$ from the fit. In fact, in our tests, we find that the removal of these bins yield a significantly better fit in terms of $\chi^2$/d.o.f. 
The covariance matrix is calculated from jackknife sub-samples and is described in detail in Section~3.1 of \citet{2021Yuan}.

\subsection{CMASS $w_p$ fit}
\label{subsec:wpfit}

To fit the observed CMASS projected galaxy 2PCF $w_p$, we start with our \textsc{AbacusSummit} base box at Planck cosmology. For this analysis, we use the cleaned \textsc{CompaSO} halos and their corresponding particles at the $z = 0.5$ snapshot. 


\begin{table}
\centering
\begin{tabular}{ c | c c c c}
\hline 
Parameter name & $\mu_{\mathrm{prior}}$ & $\sigma_{\mathrm{prior}}$ & best-fit & posterior median\\ 
\hline
$\log_{10}(M_{\mathrm{cut}}/h^{-1}M_\odot)$ & 13.3 & 0.5 & 12.9 & 13.1 \\ 
$\log_{10}(M_1/h^{-1}M_\odot)$ & 14.3 & 0.5 & 14.2 & 14.3 \\
$\sigma$ & 0.5 & 0.2 & 2.7$\times 10^{-3}$ & 0.26\\
$\alpha$ & 1.0 & 0.3 & 1.2 & 1.0\\
$\kappa$ & 0.5 & 0.2 & 0.08 & 0.45\\
\hline 
\end{tabular} 
\caption{The assumed priors, the maximum-likelihood values, and posterior medians of the baseline HOD model, when constrained on $w_p$. We choose the priors to be Gaussians with broad non-informative width. }
\label{tab:abacushod_bestfits}
\end{table}

\begin{figure*}
    \centering
    \hspace*{-0.6cm}
    \includegraphics[width = 5.5in]{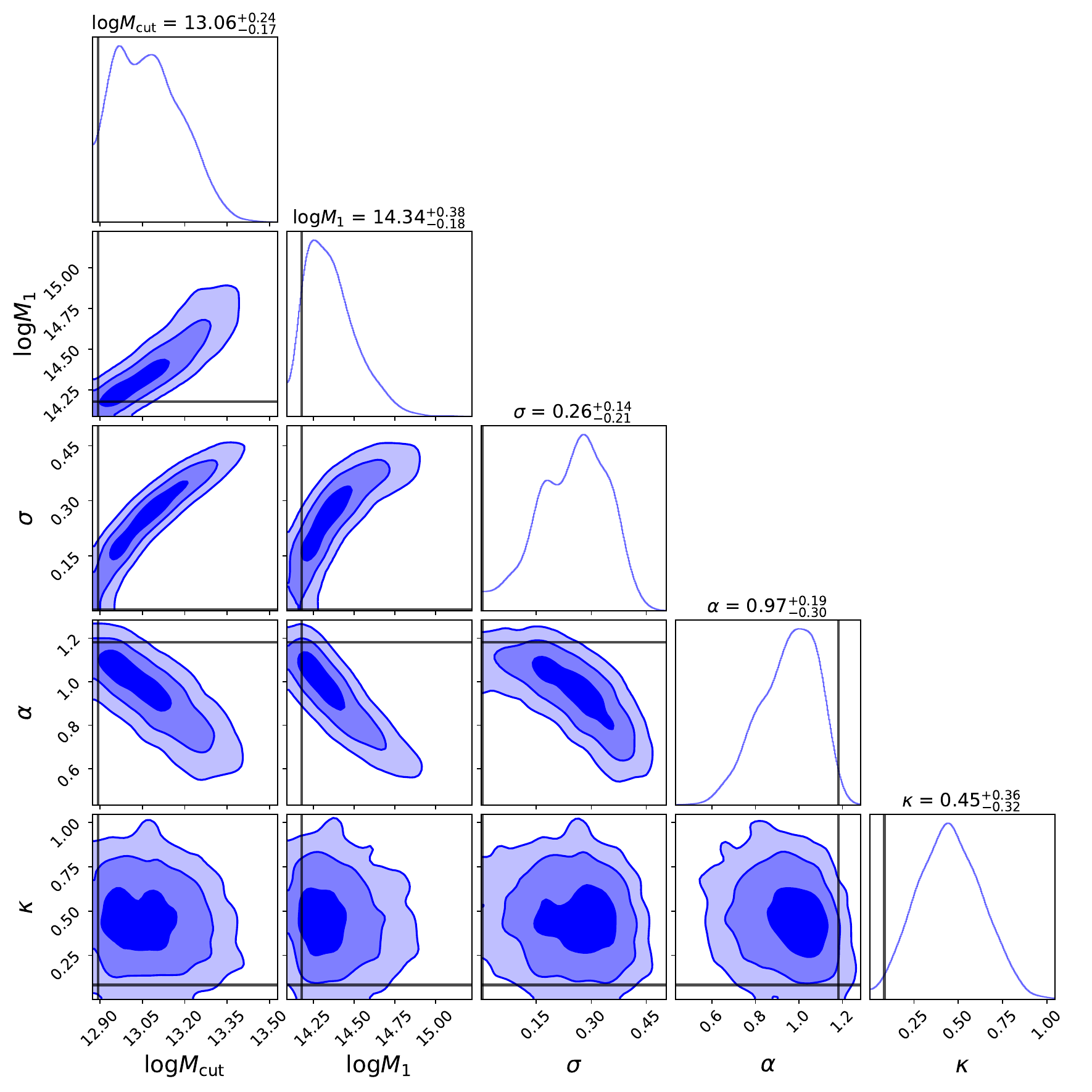}
    \vspace{-0.3cm}
    \caption{The 1D and 2D marginalized posterior constraints on the baseline HOD parameters from the $w_p$ fit. The contours shown correspond to 1, 2, and 3$\sigma$ uncertainties. The vertical and horizontal lines show the maximum-likelihood values for reference. The values displayed above the 1D marginals are posterior medians with the upper/lower bounds associated with the 0.025 and 0.975 quantiles.}
    \label{fig:corner_wp}
\end{figure*}

We assume Gaussian likelihood and express the log-likelhood in terms of the chi-squared, $\chi^2$. The $\chi^2$ is given in two parts, corresponding to errors on the projected 2PCF and errors on the galaxy number density:
\begin{equation}
\chi^2  = \chi^2_{w_p} + \chi^2_{n_g},
\label{equ:logL_wp}
\end{equation}
where
\begin{equation}
       \chi^2_{w_p}  = (w_{p,\mathrm{mock}} - w_{p,\mathrm{data}})^T \boldsymbol{C}^{-1}(w_{p,\mathrm{mock}} - w_{p,\mathrm{data}}),
       \label{equ:chi2wp}
\end{equation}
and 
\begin{equation}
   \chi^2_{n_g} = \begin{cases}
   \left(\frac{n_{\mathrm{mock}} - n_{\mathrm{data}}}{\sigma_{n}}\right)^2 & (n_{\mathrm{mock}} < n_{\mathrm{data}}) \\
   0 & (n_{\mathrm{mock}} \geq n_{\mathrm{data}}).
   \end{cases}
   \label{equ:chi2ng}
\end{equation}
where $\boldsymbol{C}$ as the jackknife covariance matrix on $\xi$, and $\sigma_n$ is the uncertainty of the galaxy number density. The $\chi^2_{n_g}$ is an asymmetric normal around the observed number density $n_\mathrm{data}$. When the mock number density is less than the data number density $(n_{\mathrm{mock}} < n_{\mathrm{data}})$, we give a Gaussian-type penalty on the difference between $(n_{\mathrm{mock}}$ and $n_{\mathrm{data}})$. When the mock number density is higher than data number density $(n_{\mathrm{mock}} \geq n_{\mathrm{data}})$, we invoke the incompleteness fraction $f_{\mathrm{ic}}$ that uniformly downsamples the mock galaxies to match the data number density. In this case, we impose no penalty. This definition of $\chi^2_{n_g}$ allows for incompleteness in the observed galaxy sample while penalizing HOD models that produce insufficient galaxy number density or too many galaxies. 
For the rest of this paper, we set $n_\mathrm{data} = 3.0\times 10^{-4} h^{3}$Mpc$^{-3}$ and a rather lenient $\sigma_n = 4.0\times 10^{-5} h^{3}$Mpc$^{-3}$.

We sample the baseline HOD parameter space, without any extensions, using the \textsc{dynesty} nested sampler \citep{2018Speagle, 2019Speagle}. While being able to sample the posterior space more efficiently than an Markov Chain Monte Carlo sampler, nested sampling codes such as \textsc{dynesty} also compute the Bayesian evidence, 
\begin{equation}
    \mathcal{Z} = P(D|M) = \int_{\Omega_{\Theta}} P(D|\Theta, M)P(\Theta|M)d\Theta.
    \label{equ:evidence}
\end{equation}
where $M$ represents the model, $D$ represents the data, and $\Theta$ represents the model parameters. The evidence can simply be interpreted as the marginal likelihood of the data given the model, and serves as an important metric in Bayesian model comparisons. Our tests show that, with sufficiently high number of live points, the nested samplings runs also are able to accurately identify the maximum likelihood point in high dimensional spaces. In our \textsc{dynesty} runs, we use 500 live points and a uniform sampler. The stopping criterion is set to $d\log\mathcal{Z} > 0.01$. We assume broad Gaussian priors for all 5 baseline HOD parameters, as summarized in Table~\ref{tab:abacushod_bestfits}.

\red{The best-fit $\chi^2 = 11$ (d.o.f = 10), and the best-fit HOD parameters are summarized in Table~\ref{tab:abacushod_bestfits}. Figure~\ref{fig:corner_wp} shows the $1,2,3\sigma$ posterior constraints. The best-fit corresponds to a galaxy number density of $n_\mathrm{fit} = 5.0\times 10^{-4} h^{3}$Mpc$^{-3}$ and a satellite fraction of 9.6$\%$. The best-fit parameters are largely within the expected range. The small $\sigma$ value corresponds to a sharp mass cut off for the central galaxies, which is reasonable given the constant mass selection cuts of CMASS galaxies. However, referring to the 2D marginalized posteriors, the typical value of $\sigma$ in the fit is closer to 0.3, and the maximum-likelihood mode might be a relative outlier.}

It is also apparent from the posterior constraints that the HOD parameters are degenerate and not well constrained. The positive correlation between $\log M_\mathrm{cut}$ and $\log M_1$ and the negative correlation between $\log M_\mathrm{cut}$ and $\alpha$ suggest a well-constrained satellite fraction. The positive correlation between $\log M_\mathrm{cut}$ and $\sigma$ suggest a well-constrained average bias on the centrals. It is possible that using average bias and satellite fraction would result in a more orthogonal HOD parameter basis. 

\begin{figure*}
    \centering
    \hspace*{-0.6cm}
    \includegraphics[width = 7in]{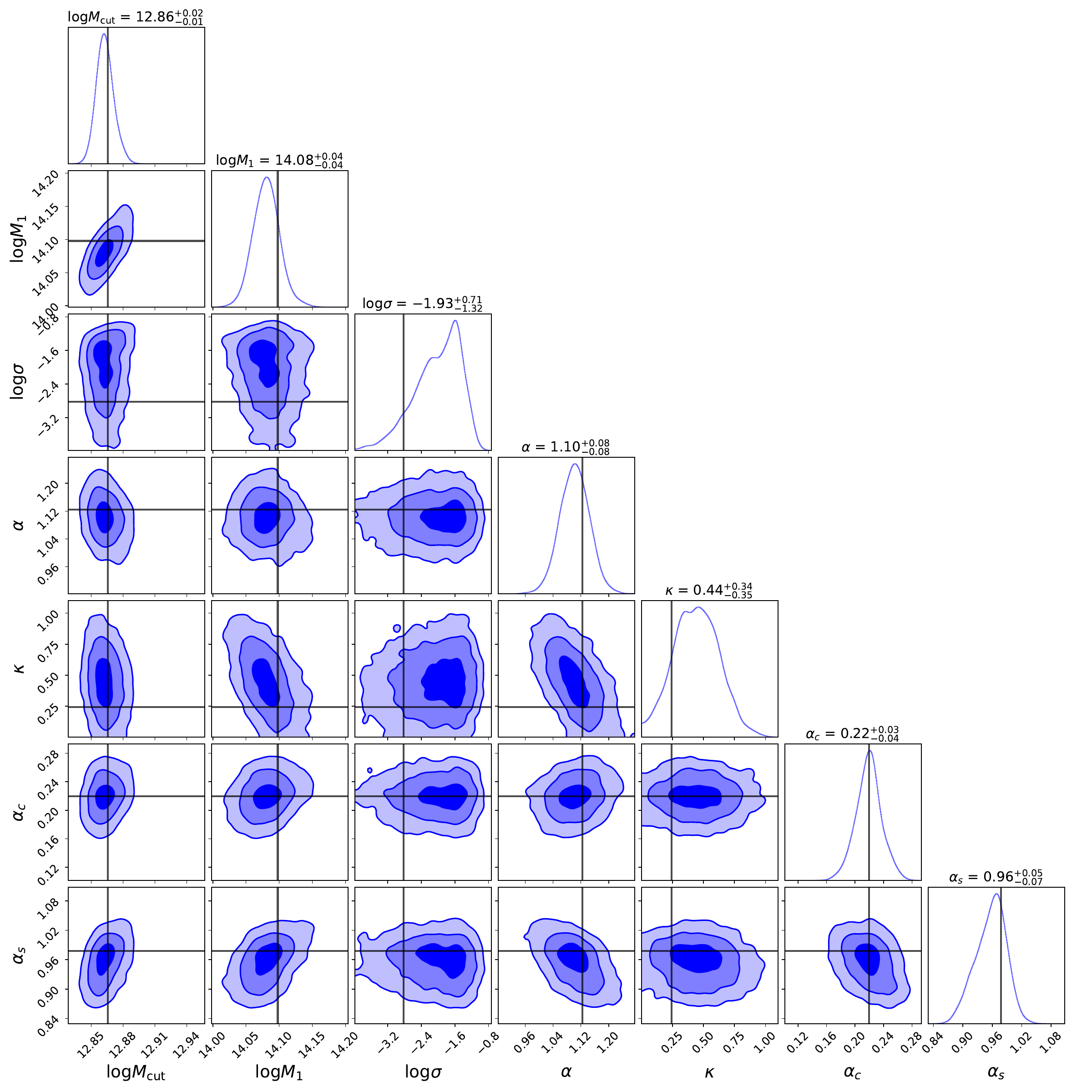}
    \vspace{-0.3cm}
    \caption{The 1D and 2D marginalized posterior constraints on the baseline HOD parameters and velocity bias paraemters from the $\xi(r_p, \pi)$ fit. The contours shown correspond to 1, 2, and 3$\sigma$ uncertainties. The vertical and horizontal lines show the maximum-likelihood values for reference. The values displayed above the 1D marginals are posterior medians with the upper/lower bounds associated with the 0.025 and 0.975 quantiles, or approximately the $2\sigma$ interval. Compared to the projected clustering $w_p$ constraints shown in Figure~\ref{fig:corner_wp}, the full-shape redshift-space clustering gives much tighter constraints on parameters and breaks multiple parameter degeneracies. }
    \label{fig:corner_xi_base_velbias}
\end{figure*}

\subsection{CMASS $\xi(r_p, \pi)$ fit}
\label{subsec:xifit}

In \citet{2021Yuan}, we found that the redshift-space 2PCF, specifically in the form of $\xi(r_p, \pi)$, offers significantly more constraining power on the HOD and assembly bias than the projected 2PCF. In this section, we present \ahod\ fits to the BOSS CMASS $\xi(r_p, \pi)$, and discuss evidence for various HOD extensions in the \ahod\ framework. 

We follow the same routine as outlined for the projected 2PCF $w_p$, using the same simulation box and the same redshift snapshot ($z = 0.5$). We also assume Gaussian likelihood, where the covariance matrix is computed from 400 jack knife samples of the data. We additionally apply corrections to the covariance matrix due to the limited simulation volume and the Hartlap correction following \citet{2007Hartlap}. Sampling was performed with \textsc{dynesty}, with the same settings as before. 

In the bare minimum, we need to extend the baseline HOD model by including velocity bias. In \citet{2021Yuan}, we used a novel physically motivated implementation of satellite velocity bias, encoded by the parameter $s_v$. Here we use the more canonical $(\alpha_s, \alpha_c)$ model of velocity bias to also account for observational systematics.  Figure~\ref{fig:corner_xi_base_velbias} showcases the posterior constraints, and the best-fit values are summarized in the third column of Table~\ref{tab:abacushod_bestfits_xi}, tagged ``Baseline'' at the top. 
 
With the $\xi(r_p, \pi)$ fit, we recover reasonable HOD parameter values. Unlike the $w_p$ fit, we do not see strong degeneracies in the marginalized posteriors. As a result, the fit yields much tighter constraints on the HOD parameters $\log M_\mathrm{cut}$, $\log M_1$, and $\alpha$. For example, the 1$\sigma$ interval on $\log M_\mathrm{cut}$ is approximately 15 times tighter than when constrained just on $w_p$. Similarly, the $1\sigma$ interval is 5 times tighter in $\log M_1$ and 3 times tighter in $\alpha$. This showcases the extra information provided in the redshift-space 2PCF. In terms of the velocity bias, we find a non-zero central velocity bias with 5$\sigma$ significance, and a satellite velocity bias consistent with 1. This measurement is consistent with the results of \citet{2015aGuo}, where the authors also found $\alpha_c \approx 0.2$ and $\alpha_s \approx 1$, albeit with lower signal-to-noise. In \citet{2021Yuan}, we found a similar $\alpha_c \approx 0.2$, but we found a significantly non-zero satellite velocity bias $s_v = 0.5-0.8$, which would suggest the satellite velocity dispersion to be larger than that of the dark matter. However, while the model difference can be partly responsible for this discrepancy, we also find the fit to be simulation dependent. \citet{2021Yuan} used the \textsc{AbacusCosmos} simulations, which are lower resolution and had a slightly different cosmology. The two simulations also use different halo finders, as detailed in \citet{2021Hadzhiyska} and \citet{2021Bose}. We find that if we fit $\xi(r_p, \pi)$ with the same HOD + $(\alpha_c, s_v)$ model as in \citet{2021Yuan}, but using the new \textsc{AbacusSummit} simulations, we recover a much smaller best-fit value $s_v \approx 0.2\pm 0.4$, statistically consistent with no satellite velocity bias. 

\begin{table*}
\centering
\begin{tabular}{ c | c c c c c}
\hline
Parameter name & $\mu_\mathrm{prior}\pm\sigma_\mathrm{prior}$ & Baseline & $B_\mathrm{cent}$, $B_\mathrm{sat}$ & $s, B_\mathrm{cent}, B_\mathrm{sat}$ & $A_\mathrm{cent}$, $A_\mathrm{sat}$ \\ 
\hline
\\[-1em]
$\log_{10}M_{\mathrm{cut}}$ & 13.3$\pm$0.5 & $12.86^{+0.01}_{-0.01}$ & $12.80^{+0.02}_{-0.02}$ & $12.78^{+0.02}_{-0.02}$ & $12.88^{+0.02}_{-0.01}$ \\ 
\\[-1em]
$\log_{10}M_1$ & 14.3$\pm$0.5 & $14.10^{+0.02}_{-0.02}$ & $14.00^{+0.04}_{-0.04}$ & $13.88^{+0.07}_{-0.05}$ & $14.17^{+0.03}_{-0.02}$\\
\\[-1em]
$\log_{10} \sigma$ & $-1\pm$1 & $-2.8^{+0.4}_{-0.7}$ & $-2.9^{+0.4}_{-0.7}$ & $-2.9^{+0.4}_{-0.7}$ & $-2.2^{+0.4}_{-0.7}$ \\
\\[-1em]
$\alpha$ & 1.0$\pm$0.3 & $1.12^{+0.04}_{-0.04}$ & $1.03^{+0.04}_{-0.04}$ & $1.05^{+0.04}_{-0.04}$ & $1.09^{+0.04}_{-0.04}$ \\
\\[-1em]
$\kappa$ & 0.5$\pm$0.2 & $0.2^{+0.2}_{-0.1}$ & $0.3^{+0.2}_{-0.2}$ & $0.5^{+0.2}_{-0.2}$ & $0.15^{+0.17}_{-0.15}$ \\
\\[-1em]
\hline
\\[-1em]
$\alpha_c$ & 0.3$\pm$0.2 & $0.22^{+0.02}_{-0.02}$ & $0.18^{+0.03}_{-0.04}$ & $0.10^{+0.04}_{-0.05}$ & $0.22^{+0.02}_{-0.02}$ \\
\\[-1em]
$\alpha_s$ & 1.0$\pm$ 0.3 & $0.98^{+0.03}_{-0.04}$ & $1.00^{+0.03}_{-0.03}$ & $0.84^{+0.07}_{-0.05}$ & $0.98^{+0.03}_{-0.03}$ \\
\\[-1em]
$s$ & 0.0$\pm$0.3 &  / & / & $-0.63^{+0.2}_{-0.1}$ & / \\
\\[-1em]
$A_\mathrm{cent}$ & 0.0$\pm$0.3 &  / & / & / & $-0.40^{+0.09}_{-0.17}$\\
\\[-1em]
$A_\mathrm{sat}$ & 0.0$\pm$0.3 & / & / & / & $0.2^{+0.2}_{-0.3}$ \\
\\[-1em]
$B_\mathrm{cent}$ & 0.0$\pm$0.3 &  / & $-0.04^{+0.02}_{-0.02}$ & $-0.04^{+0.03}_{-0.03}$ & / \\
\\[-1em]
$B_\mathrm{sat}$ & 0.0$\pm$0.3 & / & $-0.17^{+0.11}_{-0.12}$ &  $-0.15^{+0.09}_{-0.10}$ & / \\
\\[-1em]
\hline 
$f_{\mathrm{ic}}$ & / & 0.58 & 0.46 & 0.43 & 0.58 \\
$f_{\mathrm{sat}}$ & / & 0.11 & 0.13 & 0.15 & 0.11 \\
\hline 
$\chi^2$ (DoF) & / & 60 (35) & 42 (33) & 33 (32) & 54 (33)\\
$\log \mathcal{Z}$ & / & $-62$ & $-52$ & $-51$ & $-59$ \\
\hline 
\end{tabular} 
\caption{Summary of the key HOD fits on CMASS redshift-space 2PCF $\xi(r_p, \pi)$. The first column lists the HOD parameters, incompleteness factor $f_{\mathrm{ic}}$, satellite fraction $f_{\mathrm{sat}}$, the final $\chi^2$, degree-of-freedom (DoF), and the marginalized Bayesian evidence. The second and third columns show the prior constraints. The third column summarizes the best-fit parameter values of the baseline CMASS redshift-space 2PCF fit with baseline HOD + velocity bias ($\alpha_c, \alpha_s$). The following columns list the best-fit parameters when we introduce additional parameters as shown in the top row. The errors shown are $1\sigma$ marginalized errors. The fourth column shows that the addition of the environment-based secondary bias parameters substantially improves the fit and their inclusion is strongly preferred by the data. The negative $B_\mathrm{cent}$ and $B_\mathrm{sat}$ values suggest that galaxies preferentially occupy less massive galaxies in denser environments. Figure~\ref{fig:corner_B} shows the 2D posteriors of $B_\mathrm{cent}$ and $B_\mathrm{sat}$, showcasing a $>3\sigma$ detection. The next column shows that the satellite profile parameter $s$ further improves the fit, preferring a less concentrated satellite profile relative to the halo. The last column shows that the concentration-based assembly bias parameters moderately improve the fit. Figure~\ref{fig:corner_A} shows the 2D posteriors of the assembly bias parameters, showing a weaker detection at just above $2\sigma$. The best-fit values suggest that centrals preferentially occupy more concentrated (older) halos whereas satellites occupy less concentrated (younger) halos, which aligns with theory intuition. }
\label{tab:abacushod_bestfits_xi}
\end{table*}

\subsection{Introducing secondary biases}
\label{subsec:abfit}

\begin{figure}
    \centering
    \hspace*{-0.6cm}
    \includegraphics[width = 3.6in]{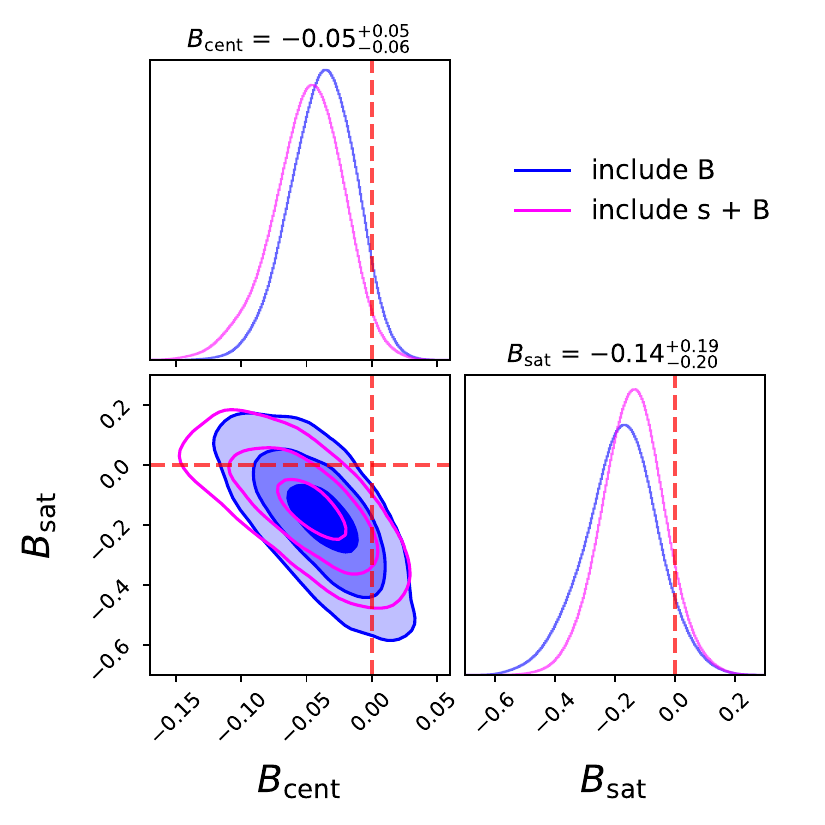}
    \vspace{-0.3cm}
    \caption{The 1D and 2D marginalized posterior constraints on the environment-based secondary bias parameters $B_\mathrm{cent}, B_\mathrm{sat}$ from the $\xi(r_p, \pi)$ fit. The contours shown correspond to 1, 2, and 3$\sigma$ uncertainties. The vertical and horizontal lines mark the zeros. The values displayed above the 1D marginals are posterior medians with the upper/lower bounds associated with the 0.025 and 0.975 quantiles, or approximately the $2\sigma$ interval. The blue contours show the constraints when we include $B_\mathrm{cent} and B_\mathrm{sat}$ in addition to the baseline HOD + velocity bias model. The magenta contours show the constraints when we also include the satellite profile parameter $s$. We see that while the marginalized 1D posteriors do not show significant detections, the 2D posterior shows that the preference for nonzero $B_\mathrm{cent}$ and $B_\mathrm{sat}$ is strong. }
    \label{fig:corner_B}
\end{figure}

\begin{figure}
    \centering
    \hspace*{-0.6cm}
    \includegraphics[width = 3.6in]{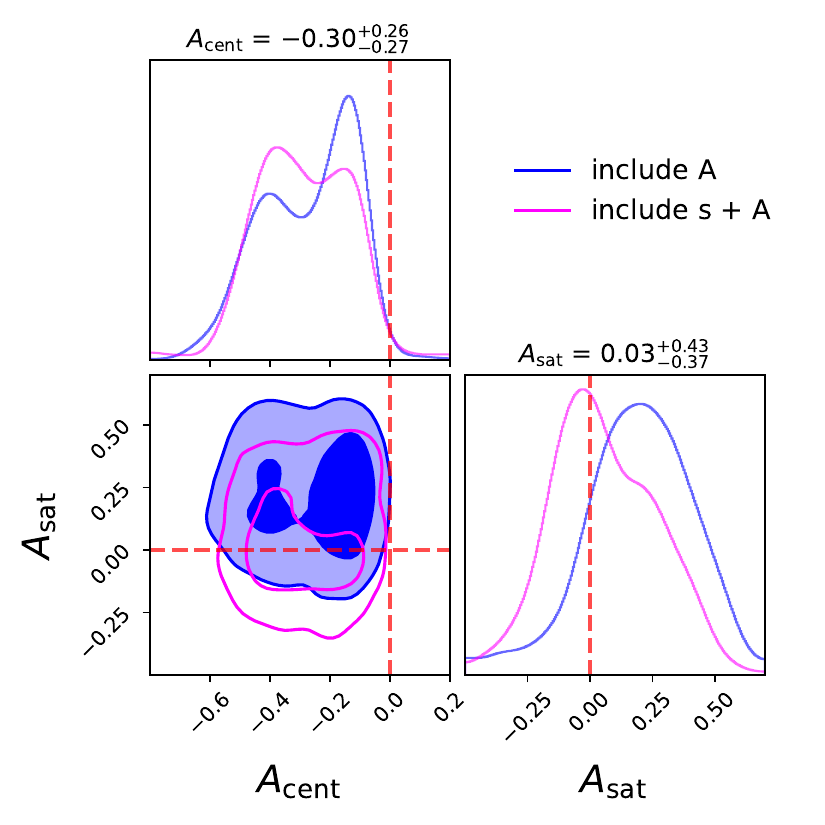}
    \vspace{-0.3cm}
    \caption{The 1D and 2D marginalized posterior constraints on the assembly bias parameters $A_\mathrm{cent}, A_\mathrm{sat}$ from the $\xi(r_p, \pi)$ fit. The contours shown correspond to 1 and 2$\sigma$ uncertainties. The 3$\sigma$ contour is less constrained, and we omit it for better visualization. The vertical and horizontal lines show the zeros. The values displayed above the 1D marginals are posterior medians with the upper/lower bounds associated with the 0.025 and 0.975 quantiles, or approximately the $2\sigma$ interval. The blue contours correspond to the constraints when only assembly bias parameters are added to the baseline HOD + velocity bias model, whereas the magenta corresponds to when we also add satellite profile parameter $s$ to the model. We find a $2\sigma$ detection of central assembly bias. The detection of satellite assembly bias is weak, especially when we include satellite profile parameter $s$. }
    \label{fig:corner_A}
\end{figure}

In this section, we further extend the baseline + velocity bias model with the environment-based secondary bias and the concentration-based assembly bias. First we extend the baseline + velocity bias model with the following parameters one at a time: $s$, $s_v$, $s_p$, ($A_\mathrm{cent}$, $A_\mathrm{sat}$), and ($B_\mathrm{cent}$, $B_\mathrm{sat}$), where each pair of secondary biases is considered one ``parameter''. We find that $s$, ($A_\mathrm{cent}$, $A_\mathrm{sat}$), and ($B_\mathrm{cent}$, $B_\mathrm{sat}$) significantly improve the best-fit $\chi^2$, with $\Delta \chi^2 = -7.1$, $\Delta \chi^2 = -6.1$, and $\Delta \chi^2 = -18.2$ respectively. The other parameters do not significantly improve the fit. The results that assembly bias and environment-based secondary bias improve the fit on the redshift-space 2PCF are qualitatively consistent with our findings in \citet{2021Yuan}. 

Since the environment-based secondary bias brings the largest improvement to the fit, we first introduce $B_\mathrm{cent}$ and $B_\mathrm{sat}$ and use \textsc{dynesty} to compute the model evidence and sample the posterior space. This model thus includes the 5 baseline HOD parameters, velocity bias ($\alpha_c, \alpha_s$), and environment-based secondary bias ($B_\mathrm{cent}, B_\mathrm{sat}$).
The fourth column of Table~\ref{tab:abacushod_bestfits_xi} summarizes the resulting best-fit parameter values. The model yields the best-fit $\chi^2 = 42$, a significant improvement over the baseline + velocity bias model. The marginalized evidence also significantly improves, suggesting that the observed redshift-space 2PCF significantly prefers the inclusion of environment-based secondary bias in the model. The blue contours in Figure~\ref{fig:corner_B} show the posterior constraints on $B_\mathrm{cent}$, and $B_\mathrm{sat}$ from this fit. We see that while the marginalized 1D posteriors do not show significant detections, the 2D posterior shows that the preference for nonzero $B_\mathrm{cent}$ and $B_\mathrm{sat}$ is quite significant. The negative values of $B_\mathrm{cent}$ and $B_\mathrm{sat}$ are consistent with the positive $A_e$ value reported in \citet{2021Yuan} due to the definitional differences. In both analyses, we find that the data preferentially put galaxies in halos in denser environments. 
It is also worth noting that we find the environment-based bias for the satellite galaxies to be stronger than that for the central galaxies. This shows the need for separate secondary bias prescriptions for the centrals and satellites, as opposed to the unified prescription used in \citet{2021Yuan}.

Revisiting the other parameters in this fit, we see that we continue to find strong evidence for central velocity bias, but no evidence for satellite velocity bias. Interestingly, the baseline HOD parameter fits seem to be sensitive to the inclusion of environment-based secondary bias. Specifically we see a decrease in $M_\mathrm{cut}$, $M_1$, and $\alpha$ compared to the fit without the secondary bias. These decreases translate to moving both centrals and satellites to lower mass halos. This is the same preference we saw in \citet{2021Yuan}, which in turns decreases the predicted weak lensing signal. We revisit the lensing discussion in the following section.

The fifth column of Table~\ref{tab:abacushod_bestfits_xi} summarizes the best-fit parameters when we also include the satellite profile parameter $s$ in addition to the environment-based secondary bias parameters ($B_\mathrm{cent}, B_\mathrm{sat}$). We see a further improvement to the best-fit $\chi^2$, down to 33 with 32 degrees of freedom. The model evidence also sees a further improvement. The introduction of $s$ does not significantly bias the best-fit values of the environment-based secondary bias parameters, but does affect the best-fit values of the baseline parameters and the velocity bias parameters. Specifically, we see a further decrease in the halo mass of central galaxies and satellite galaxies. The decrease in $M_1$ while $\alpha$ remains the same results in an increase in the inferred satellite fraction. The negative $s$ itself implies a less concentrated satellite galaxy distribution relative to the halo profile, preferring the outer regions of the halo over the halo core. The magenta contours in Figure~\ref{fig:corner_B} show that the inclusion of $s$ does not alter the posterior constraints on environment-based secondary bias parameters. 

\begin{figure*}
    \centering
    \hspace*{-0.6cm}
    \includegraphics[width = 7in]{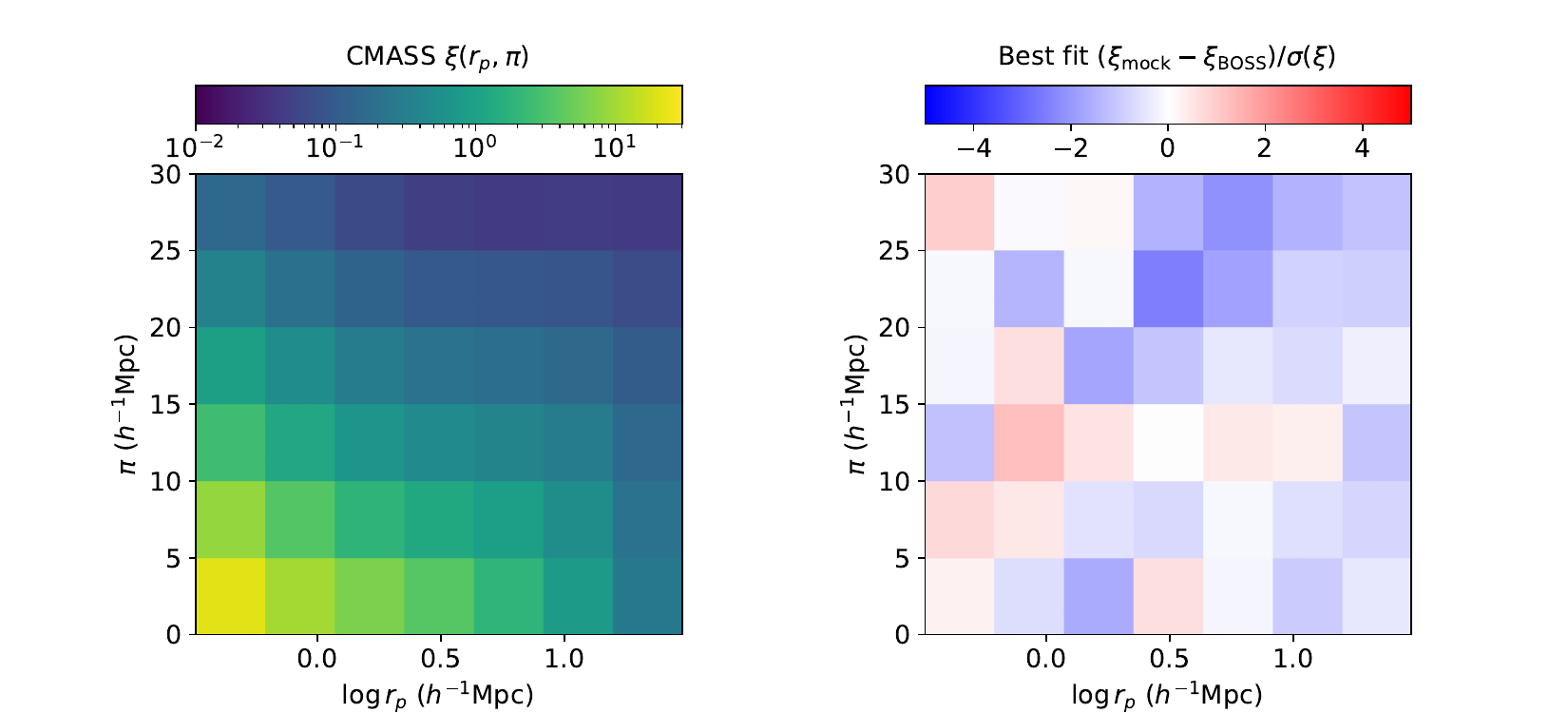}
    \vspace{-0.3cm}
    \caption{\red{The CMASS $\xi(r_p, \pi)$ that we fit to in this section (left panel) and our best fit with environment-based secondary biases (right panel). The right hand side specifically shows the difference between the best fit and the data, normalized by the data error bars, computed from the diagonal of the jackknife covariance matrix. We achieve good fit on most bins, with the exceptions being the a few bins at a few megaparsecs transverse and large $\pi$.} }
    \label{fig:xifit}
\end{figure*}

\red{
Figure~\ref{fig:xifit} showcases our best fit with environment-based secondary bias, corresponding to the fifth column of Table~\ref{tab:abacushod_bestfits_xi}. Specifically, the left hand side shows the target data vector of our analysis, i.e. the CMASS redshift-space $\xi(r_p, \pi)$ measurement. The right hand side shows the difference between our best fit and the data vector, normalized by data error bars, which we compute from the diagonal of the data covariance matrix. We achieve good fit on most bins, within $1-2\sigma$, with the exception being a few bins at $3-5h^{-1}$Mpc transverse and large $\pi$. However, note that these bins at larger $r_p$ and $\pi$ tend to be covariant, so the diagonal errors quoted underestimate the true level of uncertainty in the data, and the discrepancy between the data and the best fit is less statistically significant. There is, however, a trend the transverse direction, where the model tends to overestimate at small $r_p$ but underestimate at larger $r_p$. This suggests there is still a small residual signal that our model has not accounted for fully. }

Next, we test the baseline + velocity bias ($\alpha_c, \alpha_s$) + assembly bias ($A_\mathrm{cent}, A_\mathrm{sat}$) model. Again, we use \textsc{dynesty} to compute the model evidence and sample the posterior space. 
The sixth column of Table~\ref{tab:abacushod_bestfits_xi} summarizes the best-fit parameter values when we include parameter $A_\mathrm{cent}$ and $A_\mathrm{sat}$. 
The inclusion of the assembly bias parameters moderately reduces the $\chi^2$ per degree of freedom, and improves the marginalized Bayesian evidence.
The blue contours of Figure~\ref{fig:corner_A} show the posterior constraints on the assembly bias parameters. The 2D constraints show that the detection of central assembly bias is approximately $2\sigma$ confidence, whereas that of satellite assembly bias is less than 1$\sigma$. The magenta contours show the constraints when we also add the satellite profile parameter $s$. We recover the same constraints for the central assembly bias but no evidence for the satellite assembly bias. Overall, the evidence for assembly bias is less significant than that of environment-based secondary bias. 

The central and satellite assembly bias also seem to exhibit opposite behaviors. The best-fit values suggest that the centrals tend to live in more concentrated halos while the satellites prefer to live in less concentrated halos. This is consistent with the fact that more concentrated halos tend to be older, where more satellites have already merged with the central.
The discrepant assembly bias signature between centrals and satellites has also recently been found in the BOSS LOWZ sample \citep{2021Lange}, though they found the assembly bias signature to be dependent on redshift. It also appears, based on the 2D constraints, that the central assembly bias and the satellite assembly bias are seemingly uncorrelated. 
 



\subsection{LOWZ fits}

We repeat our analysis for the BOSS LOWZ sample in redshift range $0.15 < z < 0.40$. We continue to find that the baseline 5-parameter HOD model provides a good fit on the projected galaxy 2-point correlation function $w_p$, yielding best-fit $\chi^2 = 10$ with a DoF = 10.
We also find consistent results in the full-shape $\xi(r_p, \pi)$ fit. Most notably, the environment-based secondary bias continues to enable a significantly better fit on the observed redshift-space clustering than the concentration-based assembly bias. Specifically, without any secondary biases, the baseline HOD plus velocity bias model achieves a best-fit $\chi^2/\mathrm{DoF} = 1.42$. With assembly bias parameters $A_\mathrm{cent}$ and $A_\mathrm{sat}$, we get $\chi^2/\mathrm{DoF} = 1.36$, but with environment-based bias parameters $B_\mathrm{cent}$ and $B_\mathrm{sat}$, we get a much improved $\chi^2/\mathrm{DoF} = 0.97$. This is similar to the behavior we see for the CMASS fits in Table~\ref{tab:abacushod_bestfits_xi}. We skip the detailed figures and tables for brevity.

\section{Application to \lowercase{e}BOSS multi-tracer clustering}
\label{sec:eboss}
\red{
In this section, we apply the \ahod\ framework to fitting the multi-tracer cross-correlation measurements of the eBOSS galaxy samples. This serves as a scientifically interesting application that showcases the multi-tracer capabilities of the \ahod\ framework. 
}

\subsection{The eBOSS sample}
\red{
The dataset comes from the extended Baryon Oscillation Spectroscopic (eBOSS) survey \citep{2016Dawson}. The eBOSS project is one of the programmes within the wider 5-year Sloan Digital Sky Survey-IV \citep[SDSS-IV;][]{2017Blanton}. The eBOSS sample consists of four different types of tracers, namely LRGs; Emission Line Galaxies (ELG); Quasi-Stellar Objects (QSO); and Lyman Alpha Forest. For this analysis, we are using a subset of the eBOSS samples that covers the redshift range from 0.7 to 1.1, where all of the three tracers of interest, namely LRG, ELG and QSOs, overlap. The overlap region can be used to study these tracers with cross-correlations and this results in dense enough galaxy samples to probe the underlying dark matter distribution through the combined samples. We use intermediate versions of the data release 16 (DR16) catalogues produced by the eBOSS collaboration \citep[][]{2021Raichoor}. Any changes between the version we have used and the final versions are expected to be minor and mainly affect the results at large scales. The details of target selection and the cross-correlation measurements are found in \citet{2020Alam}. The mean number density per tracer is $n_\mathrm{LRG} = 1\times 10^{-4}h^3$Mpc$^{-3}$, $n_\mathrm{ELG} = 4\times 10^{-4}h^3$Mpc$^{-3}$, and $n_\mathrm{QSO} = 2\times 10^{-5}h^3$Mpc$^{-3}$. The exact $n(z)$ distribution can be found in figure~1 of \citet{2020Alam}.
}

\subsection{Fitting eBOSS auto/cross-correlations}
\begin{figure*}
    \centering
    \hspace*{-0.6cm}
    \includegraphics[width = 7in]{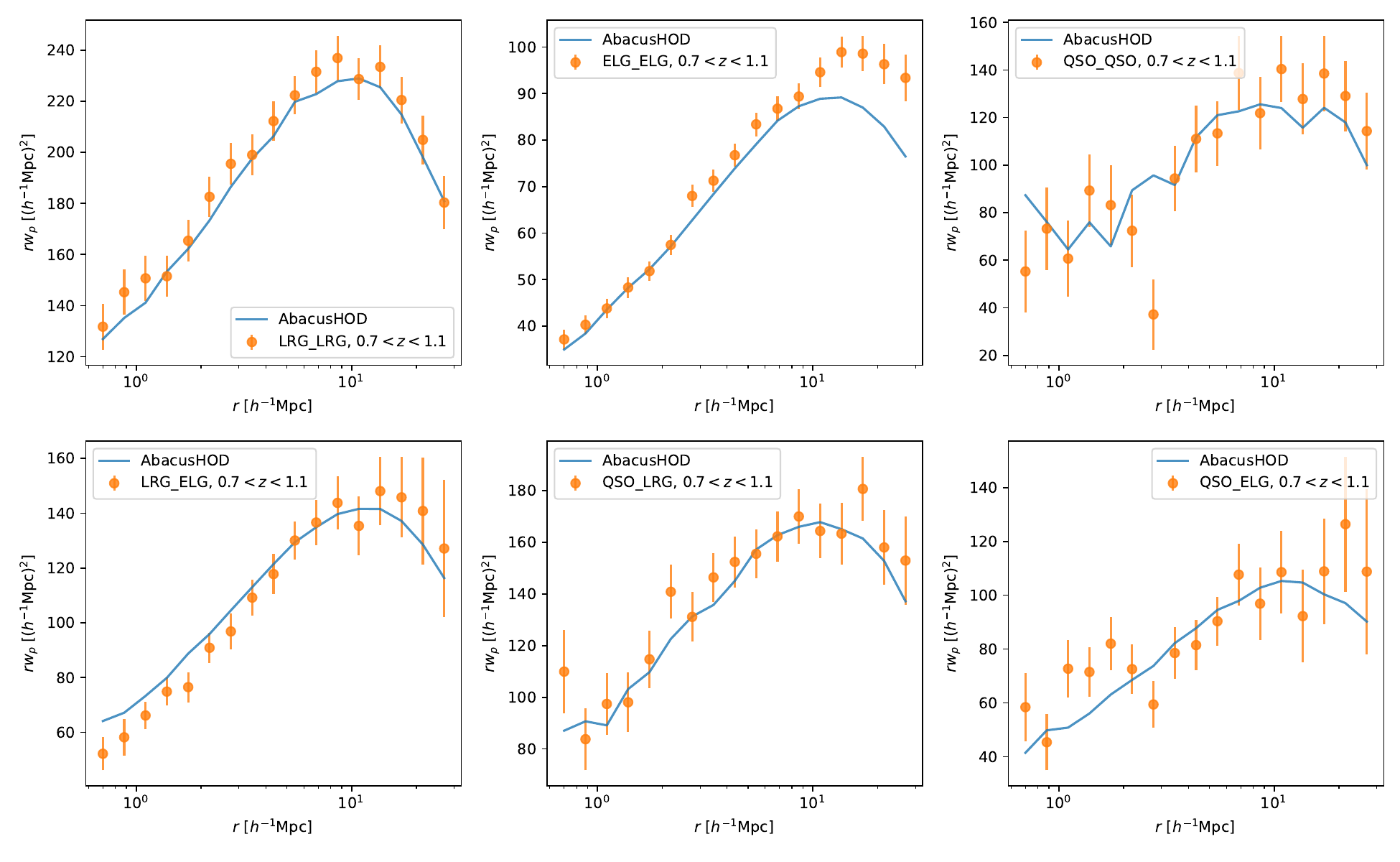}
    \vspace{-0.3cm}
    \caption{\red{The eBOSS auto/cross-correlation measurements in orange, and the \ahod\ best-fit in blue. We achieve a good fit, with the best-fit $\chi^2 = 89$ and DoF = 82. Visually, the ELG auto-correlation function appears to show large discrepancies between the data and the prediction in the largest $r$ bins. However, those bins are highly covariant, and the errorbars shown significantly underestimate the true level of uncertainty.} }
    \label{fig:ebossfit}
\end{figure*}

\red{
The eBOSS cross-correlations are measured within the overlapping footprint of the three tracers, resulting in a set of 6 projected auto/cross-correlation measurements, as showcased by the orange data points in Figure~\ref{fig:ebossfit}. The errorbars are estimated from the jackknife covariance for each measurement. We refer the readers to Section~4 of \citet{2020Alam} for a detailed description of the data vector measurements and the associated systematics.}

\red{
For this analysis, we only invoke the baseline HOD for each of the three tracers in \ahod (Equation~\ref{equ:zheng_hod_cent}-\ref{eq:alam_hod_elg}), for a total of 20 HOD parameters. Additionally, we account for incompleteness in each tracer. We define the $\chi^2$ similar to Equation~\ref{equ:logL_wp}, except the $\chi^2_{w_p}$ is now the summation of 6 individual $\chi^2$ terms, one for each auto/cross-correlation measurement, and the $\chi^2_{n_g}$ term is the summation of 3 terms, one for each tracer. We calculate a jackknife covariance for each of the auto/cross-correlation measurement, but we do not account for the covariance between the 6 measurements in this analysis. To derive the best fit, we follow the methodology of \citet{2021Yuan} in using a global optimization technique known as the covariance matrix adaptation evolution strategy \citep[CMAES;][]{2001Hansen}. We use an implementation that is part of the publicly available StochOPy (STOCHastic OPtimization for PYthon) package.\footnote{ https://github.com/keurfonluu/StochOPy}.}

\red{
We achieve a good fit on the data, with the best-fit $\chi^2 = 89$ and DoF = 82. We showcase the best-fit with the blue curves in Figure~\ref{fig:ebossfit}. Visually, the largest deviation comes from the large-scale bins of the ELG auto-correlation function. However, those large scale bins are highly covariant, and the errorbars shown significantly underestimate the true level of uncertainties. These measurements will be dramatically improved with DESI. Compared to the best fit shown in Figure~5 of \citet{2020Alam}, we see broad consistencies between the two fits. This is to be expected as the two fits use equivalent HOD models, though implemented on different simulations. Another difference is that the \citet{2020Alam} analysis only fits the 3 auto-correlation functions whereas we fit all 6 measurements simultaneously. 
}

\begin{figure}
    \centering
    \hspace*{-0.6cm}
    \includegraphics[width = 3.5in]{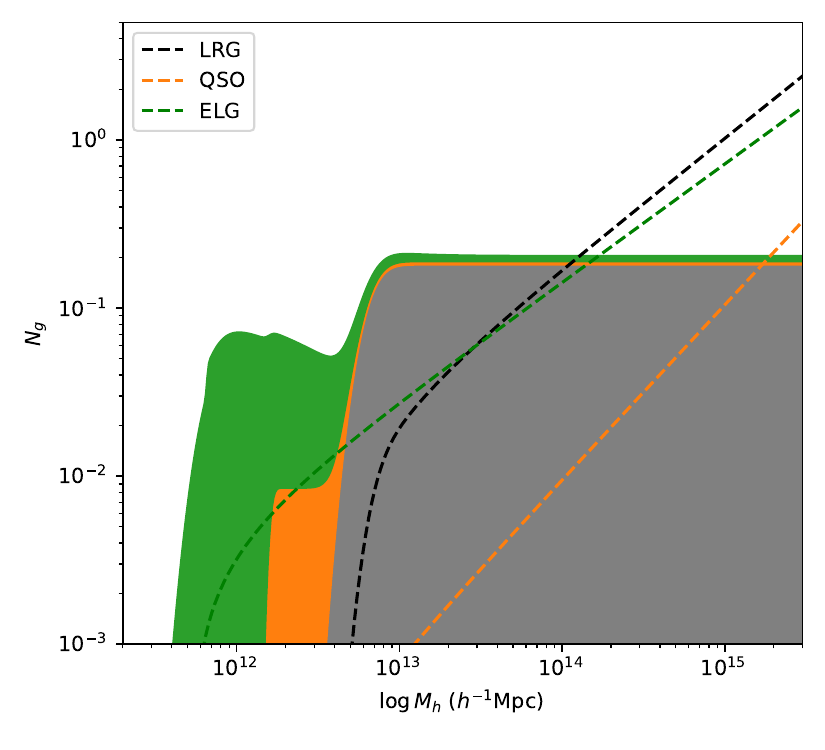}
    \vspace{-0.6cm}
    \caption{\red{The best-fit HOD model as a function of halo mass for all three eBOSS tracers. The corresponding parameter values are listed in Table~\ref{tab:ebosshod}. The shaded areas show the central occupation, stacked to show the total galaxy occupation as a function of halo mass. The dashed curves show the satellite distribution. This plot shows the different mass regimes of the three tracers, with LRGs occupying the most massive halos, whereas ELGs occupation extends down to much lower mass.}
    }
    \label{fig:mthod}
\end{figure}

\red{
The best-fit HOD parameters are summarized in Table~\ref{tab:ebosshod} and visualized in Figure~\ref{fig:mthod}. Compared to Table~1 of \citet{2020Alam}, there are some inconsistencies. However, these differences can be due to differences in HOD implementation and specifications of the simulations and halo finders. The more important point is that both models can model the auto/cross-correlation functions sufficiently well, and the best-fit predictions are consistent between the two models. We can compute the typical halo mass from the best-fit HOD parameters, yielding $M_h^{\mathrm{LRG}} = 1.9\times 10^{13}h^{-1}M_\odot$, $M_h^{\mathrm{ELG}} = 3.0\times 10^{12}h^{-1}M_\odot$, and $M_h^{\mathrm{QSO}} = 6.8\times 10^{12}h^{-1}M_\odot$. This is consistent with the findings of \citet{2020Alam}, where the authors found a mean mass per tracer of  $M_h^{\mathrm{LRG}} = 1.9\times 10^{13}h^{-1}M_\odot$, $M_h^{\mathrm{ELG}} = 2.9\times 10^{12}h^{-1}M_\odot$, and $M_h^{\mathrm{QSO}} = 5\times 10^{12}h^{-1}M_\odot$. While the mean halo mass of the LRGs and ELGs match exactly, our inferred QSO halo mass is somewhat larger than previous studies but within statistical uncertainty \citep[also refer to ][]{2017Laurent, 2017Rodriguez}. 
}

\red{
In terms of satellite fraction, we find the LRGs have a satellite fraction of $13\%$, consistent with \citet{2020Alam} but slightly lower than \citet{2017Zhai}. For QSOs, we find a satellite fraction of approximately $5\%$, consistent with previous QSO studies \citep[e.g.][]{2017Rodriguez}, but much lower than the outlier $30\%$ reported in \citet{2020Alam}. However, we do find a different mode in the likelihood surface that offers comparable $\chi^2$ and a much higher QSO satellite fraction ($\sim 34\%$). We reject that mode for rather extreme parameter values in other HOD parameters. Nevertheless, this highlights the limitation of an optimization analysis, and calls for a comprehensive posterior analysis. For ELGs, we find a satellite fraction of $7\%$, which is lower than previously reported values in the range of $12\%$ to $17\%$. Compared to \citet{2020Alam}, this is due to us finding both a higher $M_1$ and a lower $\alpha$. We suspect this is partially due to differences in the ELG HOD implementation and differences in halo finder. The \textsc{CompaSO} halos used in this analysis is known to deblend halos more aggressively than Friends-of-friends halo finders and \textsc{Rockstar} \citep[][]{2013Behroozi}, resulting in more ELGs being identified as centrals.
}

\begin{table}
	\centering
	\begin{tabular}{lccc} 
		\hline
		Parameters & LRG & QSO & ELG\\
		\hline
		$\log_{10}M_\mathrm{cut}$      & 12.8  & 12.2  & 11.83\\
		$\log_{10}\sigma$              & $-1.0$  & $-1.63$   & $-0.24$  \\
		$\gamma$                       & -     & -      & 5.8     \\
		$Q$                            & -     & -      & 19.0     \\
		$\log_{10}(M_1)$               & 14.0 & 14.0  & 14.0 \\
		$\kappa$                       & 0.63  & 0.63    & 0.82  \\
		$\alpha$                       & 0.78  & 1.04   & 0.70  \\
        $p_{\rm max}$                  & 1.0 (fixed)     & 0.85 & 0.68 \\
		\hline
	\end{tabular}
	\caption{The best fit \ahod\ parameters for all three eBOSS tracers. This should be compared to Table~1 of \citet{2020Alam}. The ELG column here should be compared to the ELG (HMQ) column in \citet{2020Alam}. We do not implement $p_{\rm max}$ for LRGs as it is redundant with the incompleteness factor in our implementation.}
	\label{tab:ebosshod}
\end{table}

\section{Discussions}
\label{sec:discuss}
\subsection{Sensitivity to environment definition}
The choice of $r_\mathrm{max} = 5h^{-1}$Mpc in our environment definition is somewhat arbitrary. It is the recommended value from \citet{2020Hadzhiyska}, in which the authors found  $r_\mathrm{max} = 5h^{-1}$Mpc to be best at capturing the secondary bias signature in hydrodynamical simulations. In \citet{2021Yuan}, we also found  $r_\mathrm{max} = 4-5h^{-1}$Mpc to be the value that yields the best fit. However, now that we are using a different simulation set and different halo finder, we again test different values of $r_\mathrm{max}$. We find no significant improvement to the fit as we vary $r_\mathrm{max}$. The lensing prediction of the best-fit HOD is also largely insensitive to $r_\mathrm{max}$. We also test the definition $r_\mathrm{max} = 5\times r_{98}$, where we remove any fixed scale from the definition. Again we find no improvement to the fit compared to the $r_\mathrm{max} = 5h^{-1}$Mpc case. 

An alternative, computationally cheaper, definition of the local environment is to use a density grid. Specifically, one calculates the local density smoothed over a Gaussian kernel at fixed grid points spanning the entire simulation box. The local density of each halo can be approximately from the overdensities at its nearest grid points, avoiding explicit pair counts. We test this environment definition with a grid size of $N = 1024^3$ and a Gaussian smoothing scale of 3$h^{-1}$Mpc. Surprisingly, we find that this grid-based environmental secondary bias yields no significant improvement over the no-environment HOD in either the best-fit $\chi^2$ or the model evidence. While we still need to explore more grid-based variations before declaring a clear preference, we highlight the high sensitivity of the fit to the details of the HOD model. For now, we continue to recommend the use enclosed neighbor mass as the local environment definition.

\subsection{Galaxy-galaxy lensing comparison}
\label{subsec:lensing}

\begin{figure}
    \centering
    \hspace*{-0.6cm}
    \includegraphics[width = 3.5in]{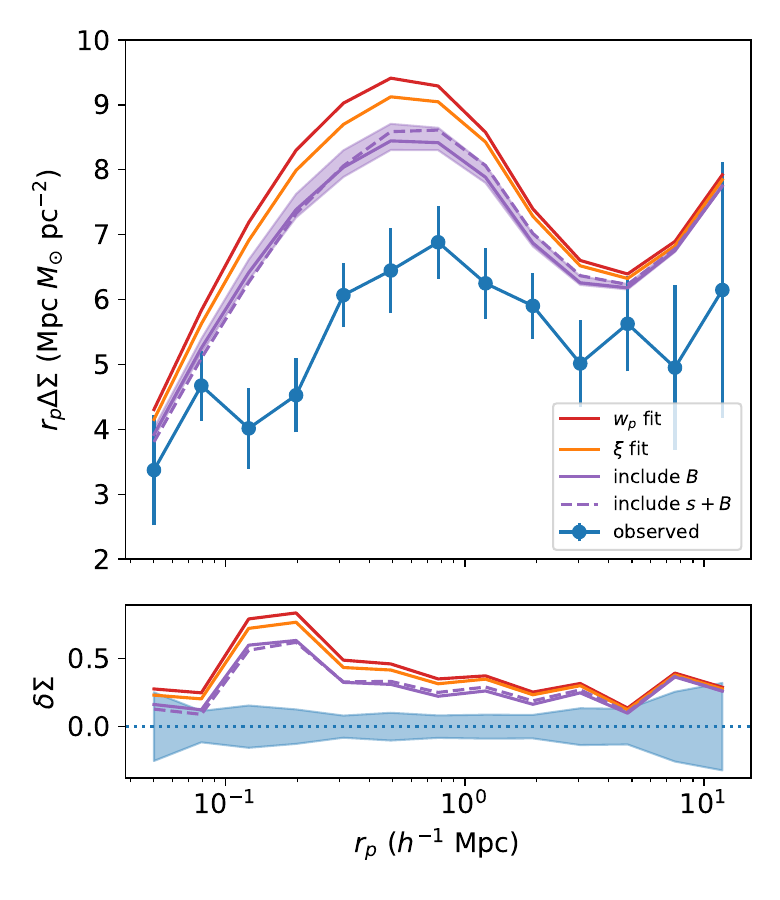}
    \vspace{-0.6cm}
    \caption{The lensing prediction of the best-fit HOD models when including the environment-based secondary bias parameters. The top panel shows the $r$ weighted surface overdensity profile, whereas the bottom panel shows the relative deviation of the predictions from the data. The red curve corresponds to the baseline HOD fit on $w_p$ presented in Section~\ref{subsec:wpfit}. The orange curve corresponds to the baseline HOD + velocity bias fit presented in Section~\ref{subsec:xifit}. The solid and dashed purple curves correspond to the environment-based secondary bias fits presented in Section~\ref{subsec:abfit}, where the dashed line also includes the satellite profile parameter $s$. The shaded purple region shows the corresponding $1\sigma$ errorbars. These fits are summarized in Table~\ref{tab:abacushod_bestfits} and Table~\ref{tab:abacushod_bestfits_xi}. 
    }
    \label{fig:lensing_B}
\end{figure}

\begin{figure}
    \centering
    \hspace*{-0.6cm}
    \includegraphics[width = 3.5in]{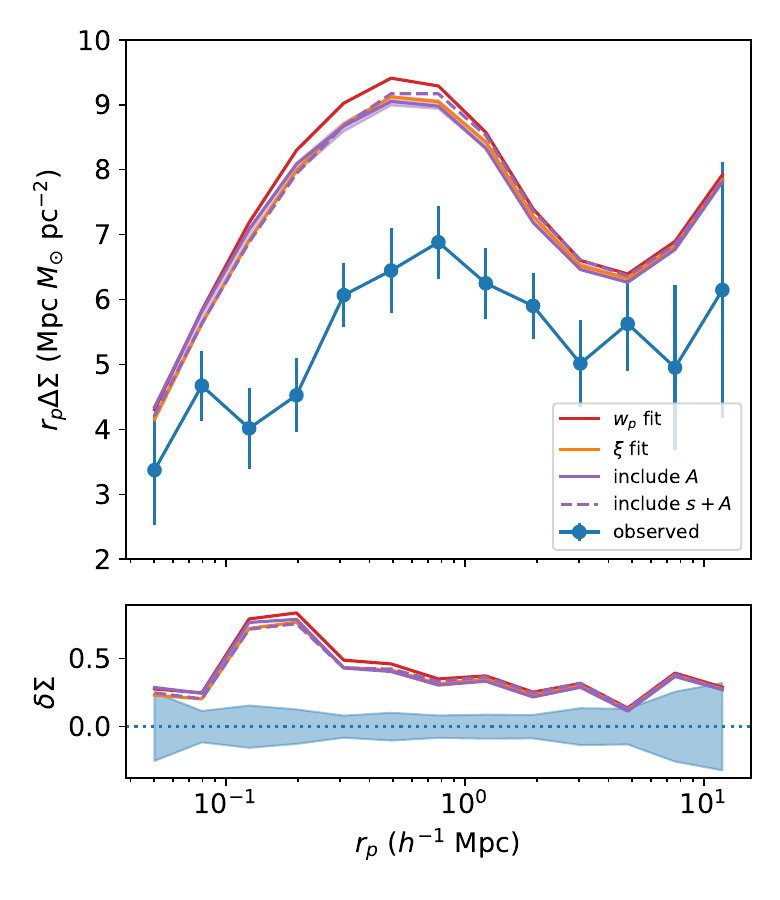}
    \vspace{-0.6cm}
    \caption{The lensing prediction of the best-fit HOD models when including the assembly bias parameters. The top panel shows the $r$ weighted surface overdensity profile, whereas the bottom panel shows the relative deviation of the predictions from the data. The red curve corresponds to the baseline HOD fit on $w_p$ presented in Section~\ref{subsec:wpfit}. The orange curve corresponds to the baseline HOD + velocity bias fit presented in Section~\ref{subsec:xifit}. The solid and dashed purple curves correspond to the assembly bias fits presented in Section~\ref{subsec:abfit}, where the dashed line also includes the satellite profile parameter $s$. The purple shaded region denotes the corresponding $1\sigma$ errorbars.}
    \label{fig:lensing_A}
\end{figure}

A well known observational tension exists between galaxy clustering and galaxy-galaxy lensing (g-g lensing). \citet{2017Leauthaud} found discrepancies of 20-40$\%$ between their measurements of g-g lensing for CMASS galaxies and a model predicted from mock galaxy catalogs generated at Planck cosmology that match the CMASS projected correlation function \citep[][see Figure~7 of \citealt{2017Leauthaud}]{2014Reid, 2016Saito}. \citet{2019Lange, 2020Lange} extended this result by finding a similar ${\sim} 25\%$ discrepancy between the projected clustering measurement and the g-g lensing measurement in the BOSS LOWZ sample. In \citet{2019Yuan}, we reaffirmed this tension by fitting simultaneously the projected galaxy clustering and g-g lensing with an extended the HOD incorporating a concentration-based assembly bias prescription. However, in \citet{2021Yuan}, we found that the inclusion of both the assembly bias and an environment-based secondary bias can significantly reduce ($\sim 10-20\%$) the predicted lensing signal when constrainted on the redshift-space 2PCF. We concluded that assembly bias and secondary biases in the HOD could be part of the explanation for the lensing tension. 

In this section, we revisit this issue with the new \textsc{AbacusHOD} fits. First, we reiterate the key differences compared to the \citet{2021Yuan} analysis. First of all, we are now using a different set of simulations, with higher spatial and force resolution, a fundamentally different halo finder, and a slight difference in cosmology. Then, we have updated the HOD model, with a different model for velocity bias, and a different model for secondary biases. Finally, while \citet{2021Yuan} used optimization routines to identify best-fits, this analysis enables full posterior sampling, achieving more robust results. 

Figure~\ref{fig:lensing_B} showcases the g-g lensing predictions of the best-fit HODs in this analysis, specifically comparing the HOD models with and without environment-based secondary biases. Again, we find that relative to the baseline HOD constrained on $w_p$, the inclusion of the environment-based secondary bias reduces the lensing prediction by $10-15\%$, towards better agreement with observation. 
This shows that the correct secondary bias models not only significantly improve the fit on the redshift-space 2PCF, but also serve an important role in resolving the g-g lensing tension. These two key results, in addition to the findings of \citet{2021Yuan}, combine to provide strong evidence for the existence of environment-based secondary bias in the CMASS galaxy population. This detection is only present in fitting the redshift-space 2PCF, demonstrating the extra information contained in the velocity space structure of the full-shape 2PCF. 

The blue ``observed'' lensing data comes from the Canada France Hawaii Telescope Lensing Survey \citep[CFHTLenS,][]{2012Heymans, 2013Miller}, and the Canada France Hawaii Telescope Stripe 82 Survey \citep[CS82,][]{2013Erben}. 
We have also internally compared these measurements to updated measurements from Hyper-Suprime Cam (HSC) survey \citep[][]{2018aMandelbaum, 2018bMandelbaum}, the Dark Energy Survey \citep[DES,][]{2016DES, 2018Drlica} Y1 and the Kilo-Degree Survey \citep[KiDS-1000,][]{2019Kuijken, 2020Wright, 2021Hildebrandt, 2021Giblin}. 
We find that the updated measurement is largely consistent with the older CFHTLens data.
The detailed comparison will be presented in upcoming papers (Amon et al. in prep, Lange et al. in prep). 

Figure~\ref{fig:lensing_A} showcases the g-g lensing predictions when including the concentration-based assembly bias parameters instead of the environment-based parameters. It is clear that the inclusion of assembly bias parameter does not help resolve the lensing tension. This is contrary to the results of \citet{2021Yuan}, where we found both assembly bias and environment-based secondary bias reduce the lensing tension. We attribute this to differences in halo finders and the assembly bias models. We discuss these differences more in Appendix~\ref{sec:cosmos}. 

\begin{figure}
    \centering
    \hspace*{-0.6cm}
    \includegraphics[width = 3.45in]{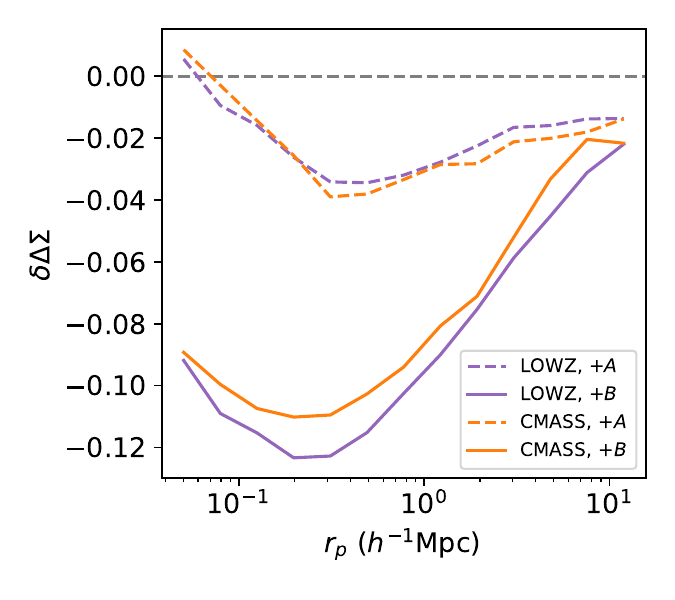}
    \vspace{-0.6cm}
    \caption{The correction to the baseline lensing prediction due to the inclusion of assembly bias (dashed lines) and environment-based bias (solid line). CMASS fits are shown in orange whereas LOWZ fits are in purple. $\delta\Delta\Sigma = (\Delta\Sigma_{\mathrm{with\ bias}}/\Delta\Sigma_\mathrm{base})-1$, where the baseline prediction comes from the naive 5-parameter HOD fit on $w_p$. We see environment-based bias consistently lowers the lensing prediction for both samples by 10$\%$ on small scales whereas assembly bias has less impact. }
    \label{fig:lensing_lowz}
\end{figure}

In LOWZ, we continue to find that the inclusion of the environment-based bias in the HOD model results in a more realistic lensing prediction, as we show in Figure~\ref{fig:lensing_lowz}. The figure shows the relative change to the predicted lensing signal due to the inclusion of assembly bias parameters (dashed lines) and environment-based secondary bias parameters (solid lines). Specifically, $\delta\Delta\Sigma = (\Delta\Sigma_{\mathrm{with\  bias}}/\Delta\Sigma_\mathrm{base})-1$, where the baseline prediction comes from the $w_p$ fit with the vanilla 5-parameter HOD model. The CMASS prediction is plotted in orange whereas the LOWZ prediction is plotted in purple. Clearly, on small scales, the $10\%$ reduction in the lensing prediction due to the environment-based bias persists over both samples.The effect of the assembly bias is also consistent across both samples, but remains small in amplitude. This lends further weight to the conclusion from the previous sections that the inclusion of environment-based biases in the HOD model not only achieves a good fit on the small-scale full shape clustering, but also reduces the lensing tension on small scales. 

We do caution that the lensing tension remains despite accounting for the environment-based secondary bias. A full resolution of the lensing tension likely requires a combination of secondary bias effects, as demonstrated here, improvements in baryonic effects modeling \citep{2020Amodeo}, and potentially a better accounting of observational systematics (Amon et al. in prep.). \red{A statistically rigorous joint-analysis of the clustering and lensing measurements is required to determine whether the combination of these effects can indeed resolve the lensing tension. We reserve such analysis to future papers. }

\subsection{Synergies with other works}
This work provides growing evidence that the local environment is an important tracer for secondary galaxy bias. \citet{2020Hadzhiyska} and \citet{2020Xu} both systematically tested the effectiveness of various secondary HOD dependencies in capturing the secondary bias signature, \citet{2020Hadzhiyska} through hydrodynamical simulations and \citet{2020Xu} through semi-analytical models. Both studies found the halo environment to be the best indicator of secondary bias. \citet{2021Yuan} and this work complement the previous works by providing the observational support for including environment-based secondary bias in HOD models.

This work also adds another piece to the ``lensing is low'' puzzle, by showing that the environment-based secondary bias can account for $30\%$ of the lensing discrepancy. Another recent development in resolving this discrepancy comes from the kinetic Sunyaev-Zeldovich (kSZ) effect measurements from the Atacama Cosmological Telescope (ACT) collaboration \citep{2020Amodeo, 2021Schaan}. These studies found that, due to baryonic feedback, the gas profile is significantly more extended in and around the host halos. A first order estimate by the ACT team shows that this extended gas profile can reduce the predicted lensing signal by approximately $50\%$. This shows that a combination of baryonic effects and secondary biases, and potentially a more thorough accounting of data systematics, can reconcile the lensing tension, without the need to invoke any change to the underlying cosmology. 

\section{Conclusions}
\label{sec:conclude}
In this paper, we present \textsc{AbacusHOD}, a new extended multi-tracer HOD framework built on top of the state-of-art \textsc{AbacusSummit} simulation suite. This HOD framework is feature rich, incorporating flexible models for secondary biases and satellite properties. The code is highly optimized for fast HOD evaluation, enabling robust sampling of extended HOD parameter spaces. In the age of DESI, this code will be an important tool in multi-tracer analyses and cosmology inference on relatively small scales. 

\red{
We present two examples applying \textsc{AbacusHOD} and \textsc{AbacusSummit} to BOSS and eBOSS data. First we model the full-shape redshift-space 2PCF on small to intermediate scales. We find that the redshift-space 2PCF is significantly more constraining on the HOD parameters than the projected 2PCF, while also breaking parameter degeneracies between $M_\mathrm{cut}, M_1, \sigma,$ and $\alpha$. 
We tested various extensions to the baseline + velocity bias model. We find that the observed redshift-space 2PCF strongly prefers the inclusion of environment-based secondary bias, with greater than $3\sigma$ detection for the environment-based secondary bias parameters. We find weaker evidence for the canonical concentration-based assembly bias, with just over $2\sigma$ detection. This is consistent with several recent studies that have found the local environment to be the far better indicator of galaxy secondary biases in the HOD.
}

\red{
In the second example, we showcase the multi-tracer capabilities of \ahod\ package by analyzing the auto/cross-correlation functions of eBOSS LRG, ELG, and QSO samples. Our model achieves a good fit on the full data vector, yielding consistent predictions with previous analyses. }

In the discussion section, we also highlight the fact that by including the environment-based secondary bias, the best-fit g-g lensing prediction is decreased by approximately 10$\%$ is magnitude, accounting for about 1/3 of the lensing tension. We also find that assembly bias does not significantly lower the lensing prediction. This result is reproduced in both the CMASS and LOWZ sample. In the general sense, this is consistent with the conclusion of \citet{2021Yuan}, that secondary biases can potentially partially resolve the lensing tension. Combined with baryonic effects, which have recently been shown to account for up to $50\%$ of the lensing tension, and a more careful accounting of data systematics, we could potentially explain the full ``lensing is low'' tension.

\section*{Acknowledgments}

We would like to thank Shadab Alam, Johannes Lange, and Josh Speagle for technical guidance in the analyses. We would like to thank David Weinberg, Charlie Conroy, Lars Hernquist, Douglas Finkbeiner for their helpful feedback.

This work was supported by U.S. Department of Energy grant DE-SC0013718, NASA ROSES grant 12-EUCLID12-0004, NSF PHY-2019786, and the Simons Foundation.
SB is supported by the UK Research and Innovation (UKRI) Future Leaders Fellowship [grant number MR/V023381/1].

This work used resources of the National Energy Research Scientific Computing Center (NERSC), a U.S. Department of Energy Office of Science User Facility located at Lawrence Berkeley National Laboratory, operated under Contract No. DE-AC02-05CH11231.
The {\sc AbacusSummit} simulations were conducted at the Oak Ridge Leadership Computing Facility, which is a DOE Office of Science User Facility supported under Contract DE-AC05-00OR22725, through support from projects AST135 and AST145, the latter through the Department of Energy ALCC program.

\section*{Data Availability}

The simulation data are available at \url{https://abacussummit.readthedocs.io/en/latest/}. The \ahod\ code package is publicly available as a part of the \textsc{abacusutils} package at \url{http://https://github.com/abacusorg/abacusutils}. Example usage can be found at \url{https://abacusutils.readthedocs.io/en/latest/hod.html}.



\bibliographystyle{mnras}
\bibliography{biblio} 




\appendix

\section{Lensing comparison to \textsc{AbacusCosmos}}
\label{sec:cosmos}
\begin{figure}
    \centering
    \hspace*{-0.6cm}
    \includegraphics[width = 3.5in]{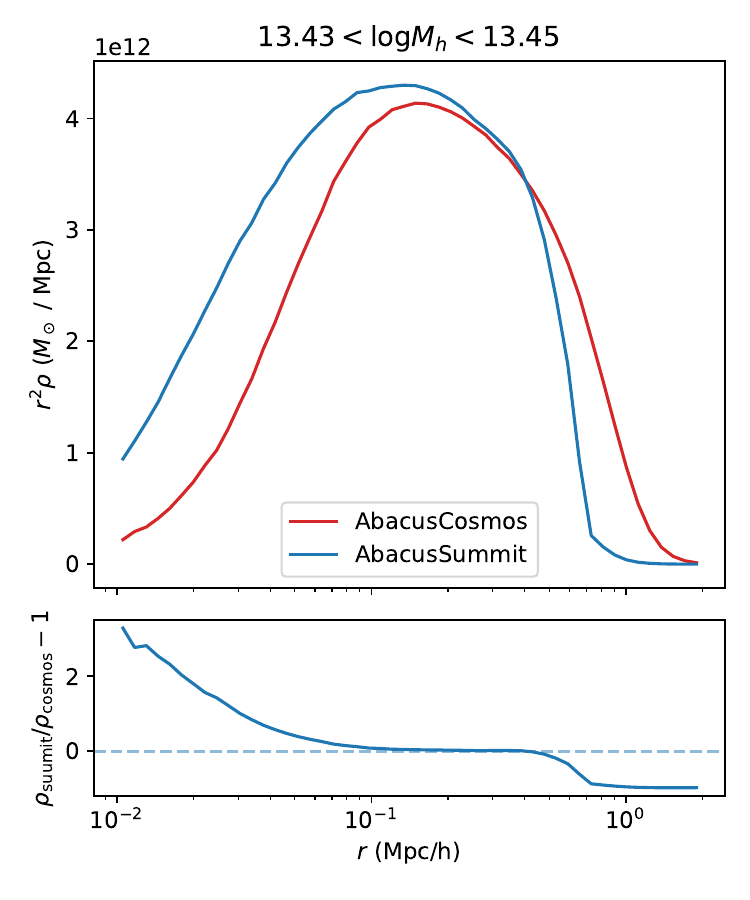}
    \vspace{-0.3cm}
    \caption{The radial density profile of two samples of halos abundance matched between \textsc{AbacusCosmos} and \textsc{AbacusSummit}. The \textsc{AbacusSummit} sample is selected to be within mass range $13.43 < \log M_h < 13.45$, which is approximately the typical mass of galaxy hosting halos. The bottom panel shows the relative difference between the two profiles, and it is apparent that the \textsc{AbacusSummit} halos have a much more concentrated core at $r < 0.1 h^{-1}$Mpc.}
    \label{fig:matched_profile}
\end{figure}
Comparing to the lensing predictions using the \textsc{AbacusCosmos} simulations presented in \citet{2021Yuan}, our updated lensing predictions are systematically higher, especially at inner halo scales ($r_p < 0.1h^{-1}$Mpc). The \textsc{AbacusCosmos} predictions have a much steeper drop off at small scales. We find that this is at least partially due to the more concentrated core structure of the \textsc{AbacusSummit} halos, which is a result of the higher force resolution in the simulation and potentially the \textsc{CompaSO} finder preferentially selecting more compact objects at fixed mass. Figure~\ref{fig:matched_profile} showcases the halo density profile of two samples of halos abundance matched between \textsc{AbacusCosmos} and \textsc{AbacusSummit}. It is clear that the \textsc{AbacusSummit} halos have a significantly more concentrated core at $r < 0.1 h^{-1}$Mpc.
\begin{figure}[h]
    \centering
    \hspace*{-0.6cm}
    \includegraphics[width = 3.5in]{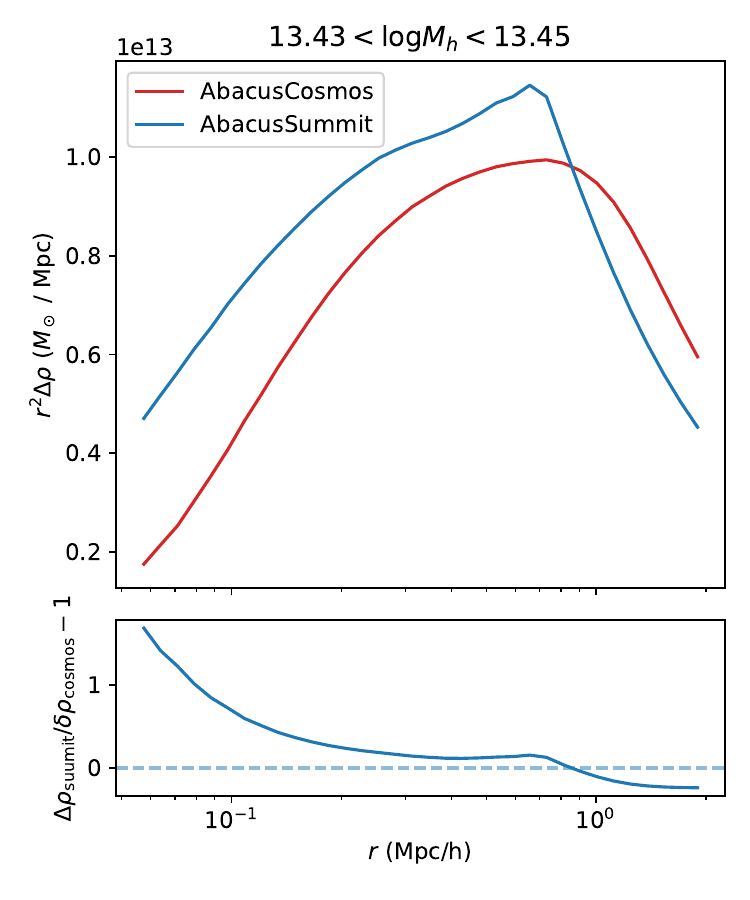}
    \vspace{-0.3cm}
    \caption{The radial overdensity profile of two samples of halos abundance matched between \textsc{AbacusCosmos} and \textsc{AbacusSummit}, the same two samples as shown in Figure~\ref{fig:matched_profile}. Note the different scale ranges in $r$ when comparing the two plots. It is clear that the difference in core structure at below $r < 0.1h^{-1}$Mpc propagates to larger scales in the overdensity. The sharp peak in the \textsc{AbacusSummit} overdensity profile at $r \approx 0.8h^{-1}$Mpc is due to deliberate choices of halo boundary in the \textsc{CompaSO} halo finder. }
    \label{fig:matched_overdensity}
\end{figure}

While the profile difference largely disappears at above 0.1$h^{-1}$Mpc, the g-g lensing measurement is a cumulative density measurement, meaning that the differences in the core propagates to larger radii. We showcase the volume overdensity $\Delta \rho$ of the same samples of abundance matched halos in Figure~\ref{fig:matched_overdensity}, where the volume overdensity is defined in a similar fashion to the lensing measurement, but in 3D:
\begin{equation}
    \Delta \rho = \bar{\rho}(<r) - \bar{\rho}(r),
    \label{equ:delta_rho}
\end{equation}
where $\bar{\rho}(<r)$ is the mean density within radius $r$, and $\bar{\rho}(r)$ is the mean density at $r$. 

It is clear that the difference in core density at $r < 0.1h^{-1}$Mpc has a significant impact on the overdensity at $r > 0.1h^{-1}$Mpc, and this effect is significant all the way up to 0.8$h^{-1}$Mpc, where the lensing signal peaks. This suggests that the sub-megaparsec $\Delta\Sigma$ measurement is likely sensitive to uncertainties on the simulated halo core structure. 

\begin{figure}
    \centering
    \hspace*{-0.6cm}
    \includegraphics[width = 3.5in]{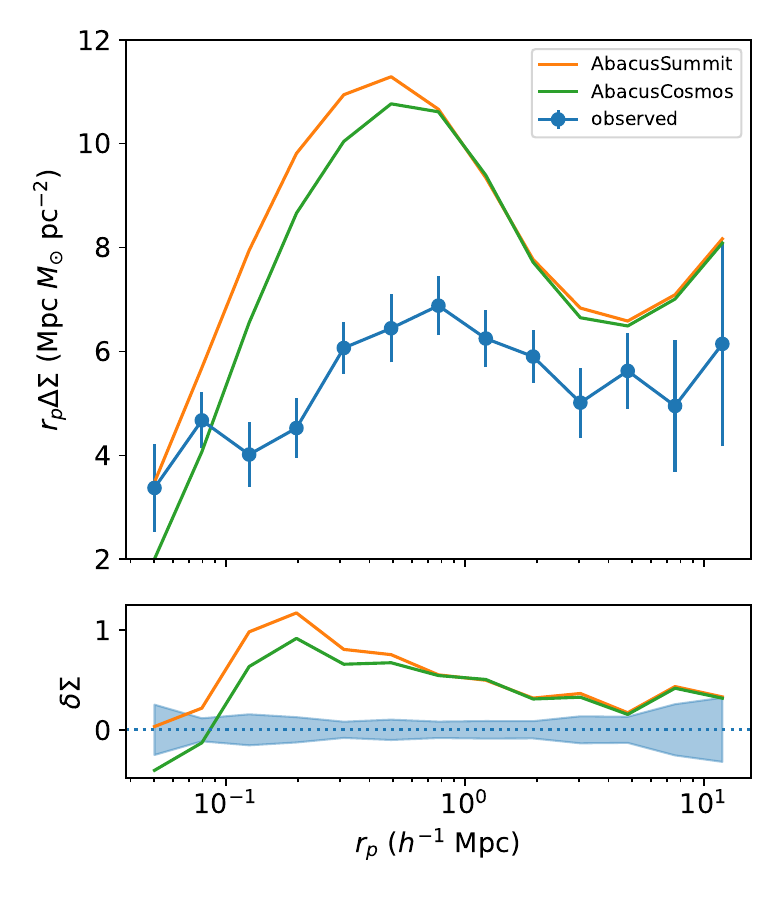}
    \vspace{-0.3cm}
    \caption{The predicted lensing measurement of halo centers abundanced matched to the BOSS CMASS mean galaxy number denisty. The bottom panel shows the relative difference between the predicted lensing measurement and the observations. The small scale discrepancy between the two simulations confirms the significant effect of halo core structure on the lensing signal.}
    \label{fig:lensing_matched}
\end{figure}

To confirm this conjecture, Figure~\ref{fig:lensing_matched} showcases the predicted lensing measurement of two abundance matched samples of halo centers, one in \textsc{AbacusCosmos} and one in \textsc{AbacusSummit}, both matched to the mean CMASS galaxy number density. 
As expected, the lensing prediction of the \textsc{AbacusCosmos} halos is biased low compared to the \textsc{AbacusSummit} halos, at scales up to 0.8$h^{-1}$Mpc. The effect is much stronger at even smaller scales of $0.1h^{-1}$Mpc. This confirms that the smallest scale lensing measurements are highly sensitive to systematics in the halo core structure. Even at moderate scales of close to $1h^{-1}$Mpc, the halo core structure remains an important systematic. Different force resolution between the \textsc{AbacusCosmos} and \textsc{AbacusSummit} simulations lead to very different core structure in equivalent halos, and this difference propagates up to megaparsec scales, at least partially explaing why the \textsc{AbacusCosmos} lensing predictions are biased low relative to the \textsc{AbacusSummit} prediction at these scales.


The other key difference with the \citet{2021Yuan} results is that we do not find significant evidence for the concentration-based assembly bias. Specifically, in the \textsc{AbacusCosmos} simulations, we find that the data prefer to put galaxies in low mass, low concentration halos, thus resulting in the strong assembly bias signature, but we do not reproduce this preference in \textsc{AbacusSummit} simulations. We believe that this is largely due to the different halo-finding algorithms and possibly different halo concentration definitions. We find that the low mass, low concentration halos in \textsc{AbacusSummit} simulations tend to be ``edge halos'' that live right on the outskirts of larger halos, resulting in a strong peak in the halo auto-correlation function at $r_p \sim 1h^{-1}$Mpc. The low mass, low concentration \textsc{AbacusCosmos} halos on the other hand, do not show this peak in the auto-correlation function, suggesting that they are a different population of halos that do not tend to live on the outskirts of large halos. This difference is a result of deliberate choices made in the different halo finders. Specifically, as shown in Figure~\ref{fig:matched_profile}, the \textsc{CompaSO} halos in \textsc{AbacusSummit} have sharper boundaries and smaller radii than \textsc{Rockstar} halos, yielding the remaining mass to additional halos. In other words, \textsc{CompaSO} deblends halos more aggressively than \textsc{Rockstar}. Neither halo-finding approaches are necessarily right or wrong, they choose to summarize the density field differently, leading to different results in assembly bias. The key is that both sets of halos can describe the data reasonably well when combined with a flexible galaxy-halo connection model.  


\bsp	
\label{lastpage}
\end{document}